\documentclass[11pt,a4paper]{article}
\pdfoutput=1
\usepackage{jheppub}
\usepackage{microtype}
\usepackage{dcolumn}
\usepackage{booktabs}

\def\bfseries{\fontseries \bfdefault \selectfont \boldmath}

\usepackage{siunitx}
\usepackage{xspace}

\newcommand{\blobone}{\textit{blob candidate 1}}
\newcommand{\blobtwo}{\textit{blob candidate 2}}

\newcommand{\bbonu}{\ensuremath{\beta\beta0\nu}}

\newcommand{\Qbb}{\ensuremath{Q_{\beta\beta}}}

\newcommand{\CS}{\ensuremath{^{137}}Cs}

\newcommand{\NA}{\ensuremath{^{22}}Na}

\newcommand{\KR}{\ensuremath{^{83m}\mathrm{Kr}}\xspace}

\newcommand{\XE}{\ensuremath{{}^{136}\rm Xe}}

\newcommand{\TL}{\ensuremath{{}^{208}\rm{Tl}}}

\newcommand{\TO}{\ensuremath{^{228}}Th}

\newcommand{\BI}{\ensuremath{{}^{214}}Bi}

\DeclareSIUnit\c{\mbox{$c$}}
\DeclareSIUnit\magn{\mbox{$\times$}}
\DeclareSIUnit\min{min}
\DeclareSIUnit\week{week}
\DeclareSIUnit\year{yr}
\DeclareSIUnit\years{years}
\DeclareSIUnit\yr{yr}
\DeclareSIUnit\standard{std}
\DeclareSIUnit\str{sr}
\DeclareSIUnit\ppm{ppm}
\DeclareSIUnit\ppb{ppb}
\DeclareSIUnit\ppt{ppt}
\DeclareSIUnit\pe{PE}
\DeclareSIUnit\spe{SPE}
\DeclareSIUnit\ev{events}
\DeclareSIUnit\ct{counts}
\DeclareSIUnit\neutron{\mbox{$n$}}
\DeclareSIUnit\smp{samples}
\DeclareSIUnit\Sample{S}
\DeclareSIUnit\ch{ch}
\DeclareSIUnit\hit{hit}
\DeclareSIUnit\hits{hits}
\DeclareSIUnit\bin{(\mbox{5-PE}~bin)}
\DeclareSIUnit\sgm{\mbox{$\sigma$}}
\DeclareSIUnit\rms{RMS}
\DeclareSIUnit\keVr{\mbox{keV$_{\rm nr}$}}
\DeclareSIUnit\keVee{\mbox{keV$_{e{\rm e}}$}}
\DeclareSIUnit\ph{photon}
\DeclareSIUnit\pes{pes}
\DeclareSIUnit\el{electrons}
\DeclareSIUnit\pm{PMT}
\DeclareSIUnit\inch{"}
\DeclareSIUnit\bit{bit}
\DeclareSIUnit\sample{samples}
\DeclareSIUnit\barn{barn}
\DeclareSIUnit\bara{bar}
\DeclareSIUnit\barg{barg}
\DeclareSIUnit\mlardepth{\mbox(meter~of~\LAr~depth)}
\DeclareSIUnit\Curie{Ci}
\DeclareSIUnit\psi{psi}
\DeclareSIUnit\parsec{pc}
\DeclareSIUnit\liveday{\mbox{live-days}}
\DeclareSIUnit\days{\mbox{days}}
\DeclareSIUnit\day{\mbox{day}}
\DeclareSIUnit\miles{\mbox{miles}}
\DeclareSIUnit\degreeC{\mbox{$^{\circ}$C}}
\DeclareSIUnit\electron{\mbox{$e^-$}}
\DeclareSIUnit\Euro{\mbox{\euro}}
\DeclareSIUnit\cph{cph}
\DeclareSIUnit\neq{neq}
\DeclareSIUnit\unit{unit}
\DeclareSIUnit\byte{Byte}
\DeclareSIUnit\Bq{\becquerel}

\begin{document}

\title{Demonstration of the event identification capabilities of the NEXT-White detector}

\collaboration{The NEXT Collaboration}
\author[15,9,a]{P.~Ferrario,\note[a]{Corresponding author.}}
\author[18]{J.M.~Benlloch-Rodr\'{i}guez,}
\author[20,15]{G.~D\'iaz L\'opez,}
\author[20]{J.A.~Hernando~Morata,}
\author[18]{M.~Kekic,}
\author[18]{J.~Renner,}
\author[18]{A.~Us\'on,}
\author[15,9,b]{J.J.~G\'omez-Cadenas,\note[b]{NEXT Co-spokesperson.}}
\author[2]{C.~Adams,}
\author[18]{V.~\'Alvarez,}
\author[6]{L.~Arazi,}
\author[19]{I.J.~Arnquist,}
\author[4]{C.D.R~Azevedo,}
\author[2]{K.~Bailey,}
\author[21]{F.~Ballester,}
\author[13]{F.I.G.M.~Borges,}
\author[3]{N.~Byrnes,}
\author[18]{S.~C\'arcel,}
\author[18]{J.V.~Carri\'on,}
\author[22]{S.~Cebri\'an,}
\author[19]{E.~Church,}
\author[13]{C.A.N.~Conde,}
\author[11]{T.~Contreras,}
\author[18]{J.~D\'iaz,}
\author[5]{M.~Diesburg,}
\author[13]{J.~Escada,}
\author[21]{R.~Esteve,}
\author[18]{R.~Felkai,}
\author[12]{A.F.M.~Fernandes,}
\author[12]{L.M.P.~Fernandes,}
\author[4]{A.L.~Ferreira,}
\author[12]{E.D.C.~Freitas,}
\author[15]{J.~Generowicz,}
\author[11]{S.~Ghosh,}
\author[8]{A.~Goldschmidt,}
\author[20]{D.~Gonz\'alez-D\'iaz,}
\author[11]{R.~Guenette,}
\author[10]{R.M.~Guti\'errez,}
\author[11]{J.~Haefner,}
\author[2]{K.~Hafidi,}
\author[1]{J.~Hauptman,}
\author[12]{C.A.O.~Henriques,}
\author[15,18]{P.~Herrero,}
\author[21]{V.~Herrero,}
\author[6,7]{Y.~Ifergan,}
\author[2]{S.~Johnston,}
\author[3]{B.J.P.~Jones,}
\author[17]{L.~Labarga,}
\author[3]{A.~Laing,}
\author[5]{P.~Lebrun,}
\author[18]{N.~L\'opez-March,}
\author[10]{M.~Losada,}
\author[12]{R.D.P.~Mano,}
\author[11]{J.~Mart\'in-Albo,}
\author[15]{A.~Mart\'inez,}
\author[18,20,c]{G.~Mart\'inez-Lema,\note[c]{Now at Weizmann Institute of Science, Israel.}}
\author[3]{A.D.~McDonald,}
\author[15]{F.~Monrabal,}
\author[12]{C.M.B.~Monteiro,}
\author[21]{F.J.~Mora,}
\author[18]{J.~Mu\~noz Vidal,}
\author[18]{P.~Novella,}
\author[3,d]{D.R.~Nygren,\note[d]{NEXT Co-spokesperson.}}
\author[18]{B.~Palmeiro,}
\author[5]{A.~Para,}
\author[23]{J.~P\'erez,}
\author[3]{F.~Psihas,}
\author[18]{M.~Querol,}
\author[2]{J.~Repond,}
\author[2]{S.~Riordan,}
\author[16]{L.~Ripoll,}
\author[10]{Y.~Rodr\'iguez Garc\'ia,}
\author[21]{J.~Rodr\'iguez,}
\author[3]{L.~Rogers,}
\author[15,23]{B.~Romeo,}
\author[18]{C.~Romo-Luque,}
\author[13]{F.P.~Santos,}
\author[12]{J.M.F. dos~Santos,}
\author[6]{A.~Sim\'on,}
\author[14,f]{C.~Sofka,\note[f]{Now at University of Texas at Austin, USA.}}
\author[18]{M.~Sorel,}
\author[14]{T.~Stiegler,}
\author[21]{J.F.~Toledo,}
\author[15]{J.~Torrent,}
\author[4]{J.F.C.A.~Veloso,}
\author[14]{R.~Webb,}
\author[6,g]{R.~Weiss-Babai,\note[g]{On leave from Soreq Nuclear Research Center, Yavneh, Israel.}}
\author[14,h]{J.T.~White,\note[h]{Deceased.}}
\author[3]{K.~Woodruff,}
\author[18]{N.~Yahlali}
\emailAdd{paola.ferrario@dipc.org}
\affiliation[1]{
Department of Physics and Astronomy, Iowa State University, 12 Physics Hall, Ames, IA 50011-3160, USA}
\affiliation[2]{
Argonne National Laboratory, Argonne, IL 60439, USA}
\affiliation[3]{
Department of Physics, University of Texas at Arlington, Arlington, TX 76019, USA}
\affiliation[4]{
Institute of Nanostructures, Nanomodelling and Nanofabrication (i3N), Universidade de Aveiro, Campus de Santiago, Aveiro, 3810-193, Portugal}
\affiliation[5]{
Fermi National Accelerator Laboratory, Batavia, IL 60510, USA}
\affiliation[6]{
Nuclear Engineering Unit, Faculty of Engineering Sciences, Ben-Gurion University of the Negev, P.O.B. 653, Beer-Sheva, 8410501, Israel}
\affiliation[7]{
Nuclear Research Center Negev, Beer-Sheva, 84190, Israel}
\affiliation[8]{
Lawrence Berkeley National Laboratory (LBNL), 1 Cyclotron Road, Berkeley, CA 94720, USA}
\affiliation[9]{
Ikerbasque, Basque Foundation for Science, Bilbao, E-48013, Spain}
\affiliation[10]{
Centro de Investigaci\'on en Ciencias B\'asicas y Aplicadas, Universidad Antonio Nari\~no, Sede Circunvalar, Carretera 3 Este No.\ 47 A-15, Bogot\'a, Colombia}
\affiliation[11]{
Department of Physics, Harvard University, Cambridge, MA 02138, USA}
\affiliation[12]{
LIBPhys, Physics Department, University of Coimbra, Rua Larga, Coimbra, 3004-516, Portugal}
\affiliation[13]{
LIP, Department of Physics, University of Coimbra, Coimbra, 3004-516, Portugal}
\affiliation[14]{
Department of Physics and Astronomy, Texas A\&M University, College Station, TX 77843-4242, USA}
\affiliation[15]{
Donostia International Physics Center (DIPC), Paseo Manuel Lardizabal, 4, Donostia-San Sebastian, E-20018, Spain}
\affiliation[16]{
Escola Polit\`ecnica Superior, Universitat de Girona, Av.~Montilivi, s/n, Girona, E-17071, Spain}
\affiliation[17]{
Departamento de F\'isica Te\'orica, Universidad Aut\'onoma de Madrid, Campus de Cantoblanco, Madrid, E-28049, Spain}
\affiliation[18]{
Instituto de F\'isica Corpuscular (IFIC), CSIC \& Universitat de Val\`encia, Calle Catedr\'atico Jos\'e Beltr\'an, 2, Paterna, E-46980, Spain}
\affiliation[19]{
Pacific Northwest National Laboratory (PNNL), Richland, WA 99352, USA}
\affiliation[20]{
Instituto Gallego de F\'isica de Altas Energ\'ias, Univ.\ de Santiago de Compostela, Campus sur, R\'ua Xos\'e Mar\'ia Su\'arez N\'u\~nez, s/n, Santiago de Compostela, E-15782, Spain}
\affiliation[21]{
Instituto de Instrumentaci\'on para Imagen Molecular (I3M), Centro Mixto CSIC - Universitat Polit\`ecnica de Val\`encia, Camino de Vera s/n, Valencia, E-46022, Spain}
\affiliation[22]{
Laboratorio de F\'isica Nuclear y Astropart\'iculas, Universidad de Zaragoza, Calle Pedro Cerbuna, 12, Zaragoza, E-50009, Spain}
\affiliation[23]{
	Laboratorio Subterr\'{a}neo de Canfranc, Paseo de los Ayerbe s/n, Canfranc Estaci\'{o}n, Huesca, E-22880, Spain}

\abstract{In experiments searching for neutrinoless double-beta decay, the possibility of identifying the two emitted electrons is a powerful tool in rejecting background events and therefore improving the overall sensitivity of the experiment. In this paper we present the first measurement of the efficiency of a cut based on the different event signatures of double and single electron tracks, using the data of the NEXT-White detector, the first detector of the NEXT experiment operating underground. Using a \TO\ calibration source to produce signal-like and background-like events with energies near 1.6 MeV, a signal efficiency of $71.6 \pm 1.5_{\textrm{ stat}} \pm 0.3_{\textrm{ sys}} \%$  for a background acceptance of $20.6 \pm 0.4_{\textrm{ stat}} \pm 0.3_{\textrm{ sys}} \%$ is found, in good agreement with Monte Carlo simulations. An extrapolation to the energy region of the neutrinoless double beta decay by means of Monte Carlo simulations is also carried out, and the results obtained show an improvement in background rejection over those obtained at lower energies.}
\maketitle

\clearpage

\section{Topological signature in \bbonu\ searches using high pressure xenon TPCs}
\label{intro}
Neutrinoless double beta decay (\bbonu) is an unobserved transition, in which two neutrons convert into protons with the emission of two electrons and no antineutrinos. The observation of this decay would imply lepton number violation and the demonstration of the Majorana nature of neutrinos. A Majorana neutrino could be one of the necessary elements to generate the matter-antimatter asymmetry of the Universe, via leptogenesis \cite{Fukugita:1986hr}.

A large experimental effort is currently ongoing to discover \bbonu\ in several isotopes and using different experimental techniques, by optimizing two main requirements, namely excellent energy resolution and background rejection. The NEXT experiment uses a high pressure time projection chamber (TPC) with electroluminescence amplification, to search for \bbonu\ in the isotope 136 of xenon.

The TPC is one of the most widely used detectors in particle physics; in the last decade, its use has grown among the experiments that search for rare events, such as dark matter or neutrinoless double beta decay \cite{Albert:2013gpz, Aprile:2018dbl}. One of the advantages of gaseous TPCs for \bbonu\, compared to liquid TPCs is that they provide a 3D image of particle tracks, which represent a useful tool to discriminate signal from background (the `topological signature'). 

\begin{figure}[htb!]
\centering
\includegraphics[width=0.49\textwidth]{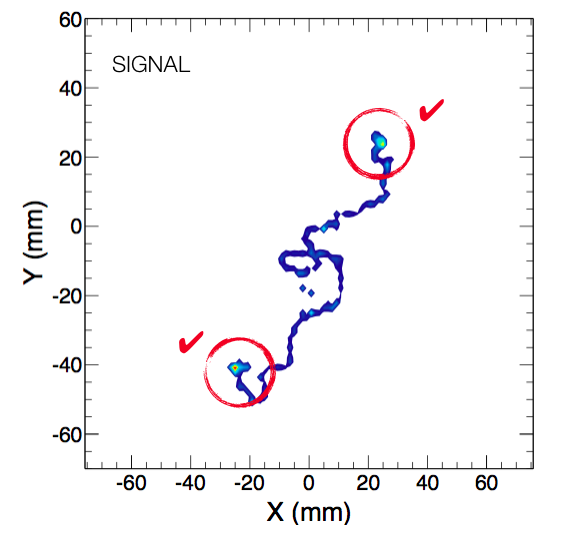}
\includegraphics[width=0.48\textwidth]{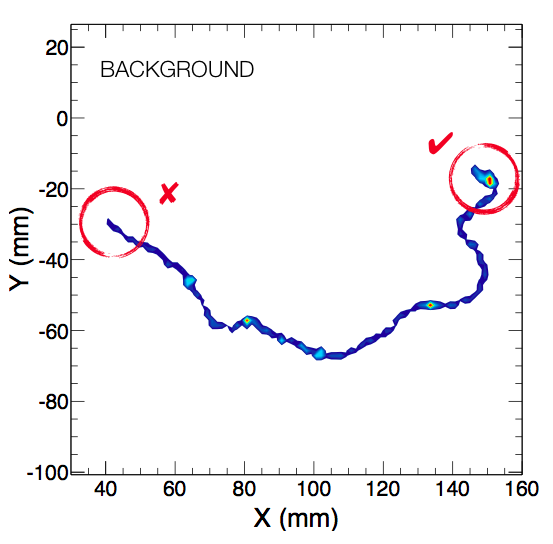}
\caption{\small A \bbonu\ event (left) and a single electron background event from a 2.447 MeV \BI\ gamma (right) in Monte Carlo simulation. Both events are simulated in gaseous xenon at 15 bar gas pressure.}
\label{fig.blobs}
\end{figure}

The signal of a \bbonu\ decay consists of two electrons originating from the same vertex, while the background comes essentially from the high energy gammas of the radioactive environment and detector components, which convert in the detection material, producing Compton and photoelectric electrons. If the energy of these electrons is close to the end-point of the $\beta\beta$ spectrum, falling within the window given by the energy resolution, they can be spuriously reconstructed as signal. However, in a gaseous TPC, signal and background can be differentiated exploiting the different patterns of their energy deposition in the gas. At a pressure of 15 bar, the two electrons emitted in a \bbonu\ decay leave a track of about 15 cm.  An electron releases its energy interacting with the gas molecules at an almost fixed rate, until the end of its range, where it produces a larger energy deposition in a smaller region, as $dE/dx \propto 1/v^2$, where $v$ is the speed of the electron, $E$ its energy and $x$ the travelled space. Therefore, the signature of a \bbonu\ event is a long track of constant energy deposition with two larger energy depositions at the end points (`blobs'), while a background event shows only one blob at one extreme of the track, as illustrated in Fig.~\ref{fig.blobs}.  
%

The first experiment that exploited the topological signature of a gaseous TPC in \bbonu\ searches was the Gotthard experiment, run by the the Caltech-Neuch\^atel-PSI Collaboration in the 1990s, which used a $\sim$ 3.3 kg \XE\ TPC at a pressure of 5 bar with multiwire read-out. It obtained an excellent signal efficiency of $68\%$ for a background rejection of single electrons of $96.5\%$, via a visual scanning of the events \cite{Luscher:1998sd}. However, the Gotthard TPC had a poor energy resolution, limited by the fluctuations in the avalanche gain and those introduced by the quenching of the scintillation light due to the gas mixture, which was used to reduce diffusion. Additionally, the quencher made the detection of the primary scintillation signal impossible, therefore preventing the z coordinate reconstruction and the rejection of background electrons coming from the cathode.

The NEXT Collaboration demonstrated a first proof of the power of the topological signature in an electroluminescent gaseous xenon TPC \cite{Ferrario:2015kta}, using the NEXT-DEMO prototype, which contained 1.5 kg of natural xenon. In that work, events in the double escape peak of the 2.6-MeV gamma coming from \TO\ decay were used to mimic the signal, while the background consisted of events in the photoelectric peak of the high energy de-excitation gamma of \NA. A signal efficiency of $66.7 \pm 0.9$ (stat.)$\pm$ 0.3 (fit)$\%$ was measured, for a background acceptance of $24.3 \pm 1.4$ (stat.)\%, in good agreement with Monte Carlo simulations. This study was limited by the small size of the NEXT-DEMO detector, in which the event selection tended to favour less extended events, with a more complicated reconstruction.
 
NEXT-White is the first stage of the NEXT-100 detector, and deploys $\sim$5 kg of xenon in an active cylindrical volume of $\sim 53$ cm of length and 40 cm of diameter, at 10 bar of pressure. Twelve photomultiplier tubes (PMTs) provide the energy measurement, while an array of 1-mm$^2$ silicon photomultipliers (SiPMs) is used for the particle track reconstruction. For a detailed description of the detector, see Ref.~\cite{Monrabal:2018xlr}. From October 2016 to early 2019, several runs of calibration and background measurements have been carried out with depleted xenon, and it has been demonstrated that an energy resolution of $\sim 1\%$ FWHM at the xenon \Qbb\ ($\sim 2458$ keV \cite{Redshaw:2007un}) can be achieved \cite{Renner:2018ttw, Renner:2}. The first run with xenon enriched in the isotope 136 has started in February 2019, with the aim of measuring the two neutrino double beta decay spectrum. 

In this work, calibration sources have been used in NEXT-White to study the performance of the topological signature to discriminate signal from background. Also, a comparison between data and Monte Carlo has been developed, in order to extrapolate the results to the \bbonu\ energy region.

The paper is organized as follows. In Sec.~\ref{reco} the particle reconstruction employed in this study is described. Section \ref{sel} explains the selection applied on data and Monte Carlo events. In Sec.~\ref{topo} the analysis procedure is presented and in Sec.~\ref{res} the results are discussed, as well as implications for the \bbonu\ region. Conclusions are drawn in Sec.~\ref{conclus}.



\section{Particle reconstruction}
\label{reco}
The electron tracks in NEXT-White can be reconstructed  by measuring the energy deposited along their path. The reference system used for the reconstruction is the natural one in a TPC, where the $z$ axis follows the drift direction, the $x$ and $y$ axes are perpendicular to the $z$ direction and the three coordinates together constitute a right-handed reference frame. Charged particles propagating in the xenon gas of the NEXT-White detector release their energy through scintillation and ionization processes. While the scintillation light (S1), detected by the PMTs, gives the starting time of the event,  the ionization charge is drifted by an electric field until it reaches the electroluminescence (EL) region, 6 mm wide, where a more intense electric field is applied and secondary scintillation (S2) is triggered. The S2 light is read both by the PMTs, which provide a precise measurement of the energy of the event, and by the SiPMs, placed $\sim$5 mm away from the EL region, which are used to reconstruct the position.
The detector triggers on the energy information read by the PMTs and provides PMT and SiPM waveforms in a buffer of a fixed size, which is always larger than the maximum possible drift time.  The sampling time of the PMTs is 25 ns, while the SiPM charge is integrated every $\mu$s. Then, the S1 and S2 signals are searched for, using the sum of the individual PMT waveforms, and the events with one S1 and one or more S2 pulses are selected for track reconstruction.

The shape of the charge pattern on the SiPMs is affected on the one hand by the longitudinal and transverse diffusion ($\sim 0.3$ and $\sim 1.1 \textrm{ mm}/\sqrt{\textrm{cm}}$ respectively \cite{Simon:2018vep}) and, on the other hand, by the spread of the light emission, which occurs along the 6 mm length of the EL region, and by the few mm distance of the SiPM plane from the emission region.  

A first cut is performed on the SiPM collected charge to eliminate dark current and electronic noise. Time bins with less than 1 photoelectron (pe) charge are suppressed, after which the total integrated charge of a SiPM is required to be above 5 pe to be considered in the reconstruction. 
These requirements have been found to eliminate most of the SiPM noise, without affecting the signal. After this first cut, the SiPM charge is rebinned to 2 $\mu$s time sections (slices) and the charge pattern is examined for each slice. For each SiPM with charge higher than 30 pe a 3D hit is generated, with $x$ and $y$ coordinates equal to the SiPM $x$ and $y$ positions and $z$ coordinate equal to the difference between the time of the slice and the time of S1, multiplied by the drift velocity of the electrons in the gas. This large charge threshold has been found to be useful to eliminate the effects of the diffusion and light spread mentioned above: it removes the charge far from the center of the source of light, keeping the information on the position of the source. The energy measured by the PMTs in the same time slice is divided among the reconstructed hits, proportionally to the charge of the SiPMs used to determine their position. If in a slice there are no SiPMs above threshold, the energy of that slice is assigned to the closest slice belonging to the same S2.

Once the hits of an event are identified, they have to be grouped into sets corresponding to different particles. To this aim, a connectivity criterium is defined, according to the following procedure. The gas volume of the detector is divided into 3D pixels (voxels) with a fixed dimension, and each voxel is given an energy equal to the sum of the energies of the hits that fall within its boundaries. The voxels that share a side, an edge or a corner are grouped into separated sets using a “Breadth First Search” (BFS) algorithm \cite{Cormen:2001:IA:580470}. These sets of voxels are regarded as the particle tracks of the event. The BFS algorithm also identifies the end-point voxels of each track, defined as the pair of voxels with the longest distance between them, where the distance between any pair of voxels is the shortest path along the track that connects them. A maximum size of the voxels is fixed, but the actual voxel size varies event by event, being optimized according to the distribution of the hits in space. This optimization tries to avoid having voxels with only one hit on a border. In this work a maximum voxel size of $1.5 \times 1.5 \times 1.5\textrm{ cm}^3$ has been used, which gives the best performance in terms of topological discrimination of signal from background. Moreover, as the distance between SiPMs is 1 cm, the current reconstruction does not allow for voxel sizes smaller than that value, since the minimum distance between hits is constrained to be also 1 cm. 

\begin{figure}[htb!]
\centering
\includegraphics[width=.49\textwidth]{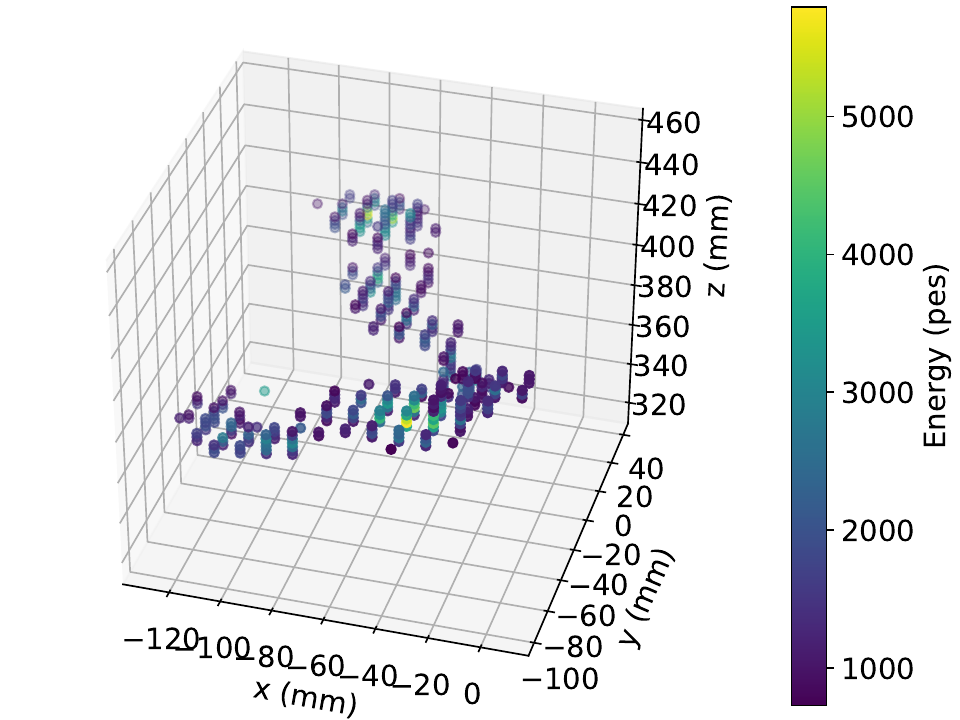}
\includegraphics[width=.49\textwidth]{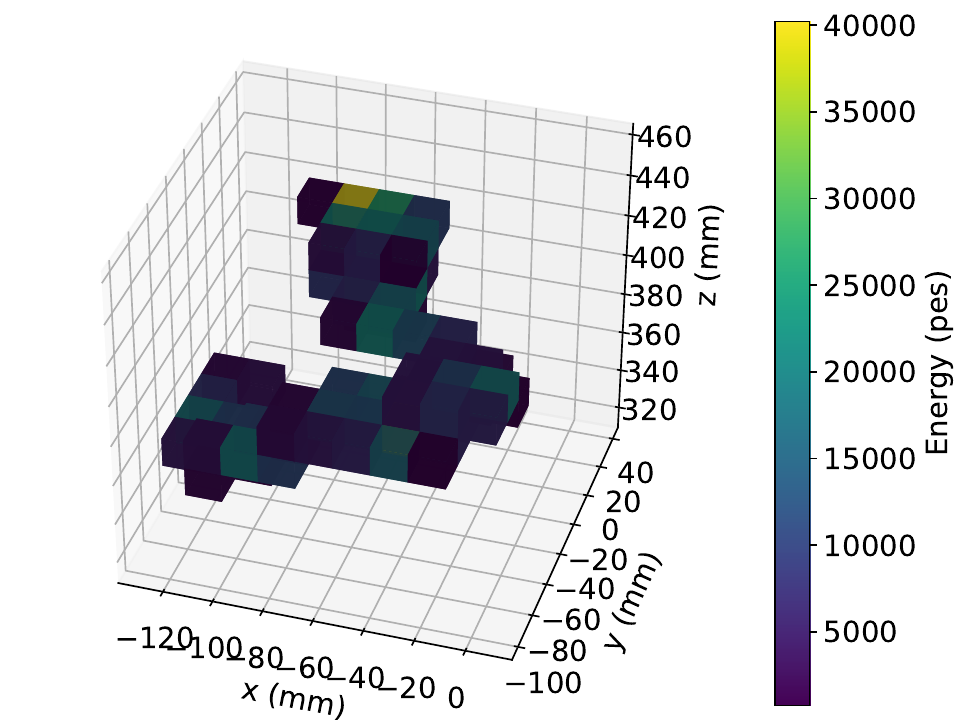}
\caption{\small Example of reconstructed hits (left) and subsequent voxelization (right). This event was produced by a \TO\ calibration source.}
\label{fig.hits}
\end{figure}

In Fig.~\ref{fig.hits}\textit{-left} the reconstructed hits of an event produced by a \TO\  calibration source are displayed. A single electron track can be seen, coming from the photoelectric interaction of the 2.615-MeV gamma. In the right plot the same reconstructed event is shown after voxelization.

\section{Data and event selection}
\label{sel}
\subsection{Data samples}

\begin{table*}[h!]\centering
\begin{tabular} {c@{\,}|c@{\,}|c@{\,}}Run number & Duration (s) & Number of triggers \\  \midrule
  6818 &   91\,248 & 525\,243 \\
  6822 & 171\,153 & 990\,892 \\
  6823 &   74\,943 & 425\,009 \\
  6826 &   93\,187 & 509\,296 \\
  6828 &   74\,233 & 432\,215 \\
  6834 &  428\,875 & 2\,495\,620 \\
 \bottomrule
\end{tabular}
\caption{Summary of the data used in this work.}\label{tab:runs}
\end{table*}

The data sets used in this work have been acquired in January 2019, during the calibration runs of the NEXT-White detector. A summary of their characteristics is presented in Table~\ref{tab:runs}. A \TO\ source was placed on the top of the detector, inserted in a feedthrough with a $z$ position in the middle of the drift region. One of the thorium daughters, \TL, decays producing a de-excitation gamma of 2.615 MeV, which can enter the active region of the detector and convert via pair production. The positron emitted in this process propagates in the gas in the same way as an electron and finally annihilates with an electron of a xenon atom, emitting two back-to-back 511-keV gammas. The energies of the electron and the positron, which are reconstructed as one track, form a peak  at 1.593 MeV in the track energy spectrum (the double escape peak) and its topology is the same as that of a \bbonu\ event, in which two electrons originate from the same point. Therefore, this peak can be exploited to study the efficiency of the reconstruction algorithms and the cuts based on the topology signature, in order to estimate their performance on the \bbonu\ signal. 
From the continuum Compton spectrum of the 2.615-MeV gamma, a sample of tracks with the same energy as the double escape peak can be extracted, and used to estimate the efficiency of background rejection.

The detector gas pressure was set to 10.1 bar and the cathode and gate voltages to 30 kV and 7.7 kV, respectively, which gave a stable drift electric field of $\simeq$0.4 kV/cm and an EL reduced electric field of $\simeq$1.27 kV/(cm$\cdot$bar). The drift velocity was very stable and it has been measured to be $\simeq$0.92 $\textrm{mm}/\mu\textrm{s}$. The electron lifetime was measured continuously using a \KR\ source diffused homogeneously in the gas and the collected charge at the PMT plane was corrected for it (for a detailed description of the NEXT-White calibration procedure, see Ref.~\cite{Martinez-Lema:2018ibw}). The average value of the electron lifetime was quite stable across the different runs, around 4.5 ms, several times larger than the full-chamber drift time. The \KR\ source provides also a map of the geometric dependence of the PMT response to EL light, which was also used to correct the detected charge for geometric effects. After these corrections, a residual dependence of the energy on the length of the track in the z-dimension was found, in which the measured energy appeared to be lower for larger tracks. A linear fit was performed to model this dependence and used to correct it. For a more detailed description of this effect, see  Ref.~\cite{Renner:2}.

The energy of the events was calibrated using a quadratic interpolation of two peaks of the \TO\ spectrum, namely the 2615-keV gamma double escape peak and photopeak, and the 662-keV photopeak of a \CS\ source placed in a lateral port. Both sources were in place at the same time and the trigger parameters were set in order to acquire both kinds of signal. In particular, the minimum charge of S2 was low enough as to include the \CS\ photopeak. As explained in Ref.~\cite{Renner:2}, a non-linearity is observed in the energy reconstruction of the events in the NEXT-White detector, therefore a linear fit to the three peaks does not produce satisfactory results. 

%

A complete Monte Carlo simulation of the decay of a \TL\ source in the same conditions as the real detector was produced, to be compared with data. The particle propagation and their energy deposition in the detector are simulated using the \texttt{nexus} software  \cite{Martin-Albo:2015dza}, a simulation package based on \textsc{Geant4} \cite{Agostinelli:2002hh}. Version \texttt{geant4.10.2.p01} has been used, together with the  \texttt{G4EmStandardPhysics\_option4} physics list. Subsequently, electron diffusion and attachment, generation of S1 and S2 light signal and their detection by PMTs and SiPMs are simulated, together with a full electronics response, using the IC framework, a simulation and reconstruction package based on python and developed by the NEXT Collaboration. The outcome of the simulation is a set of waveforms, as for data, which passes through the same reconstruction procedure described in Sec.~\ref{reco}.

\subsection{Event selection}\label{sec.ev.sel}

\begin{figure}[htb!]
\centering
\includegraphics[width=0.49\textwidth]{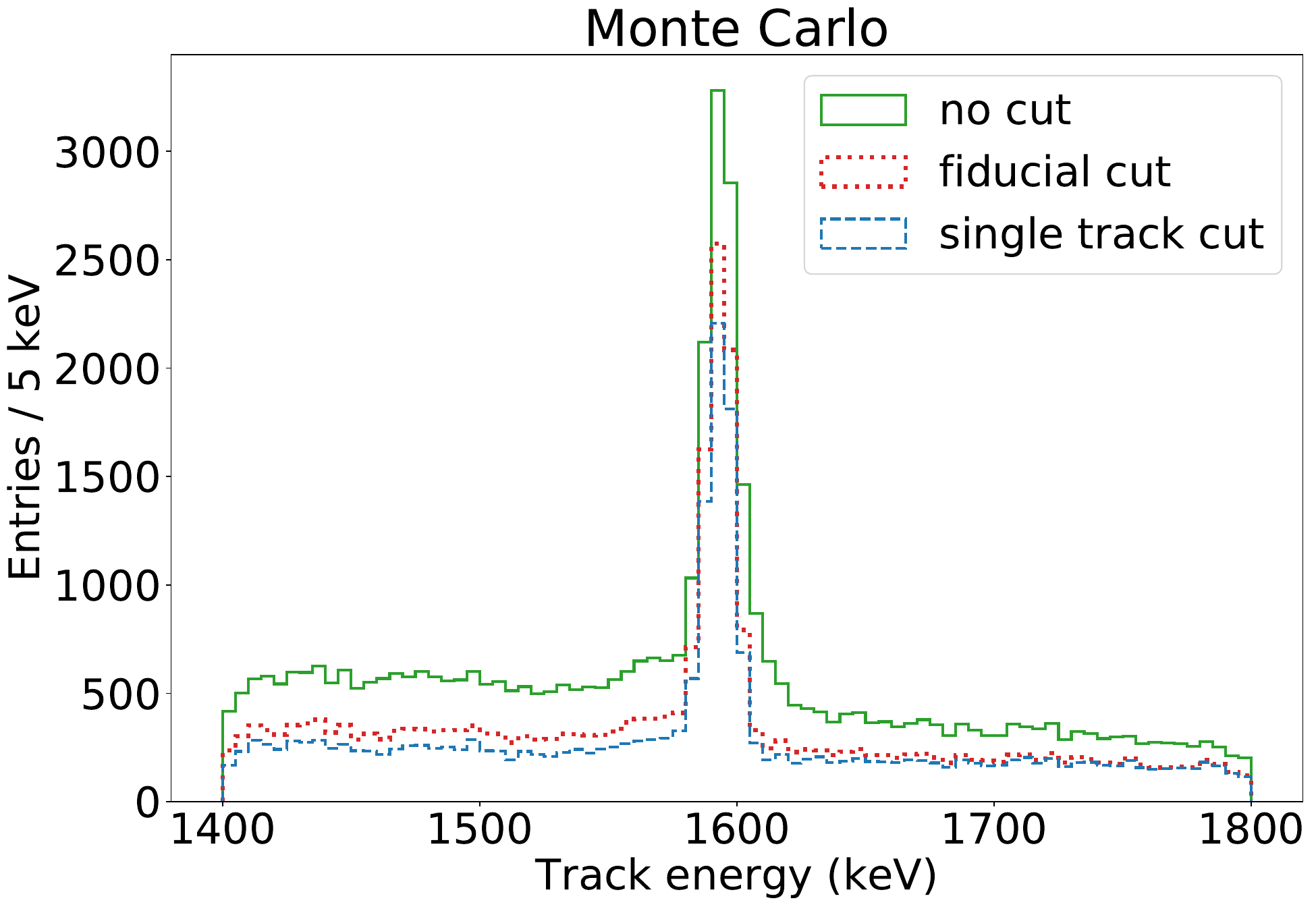}
\includegraphics[width=0.49\textwidth]{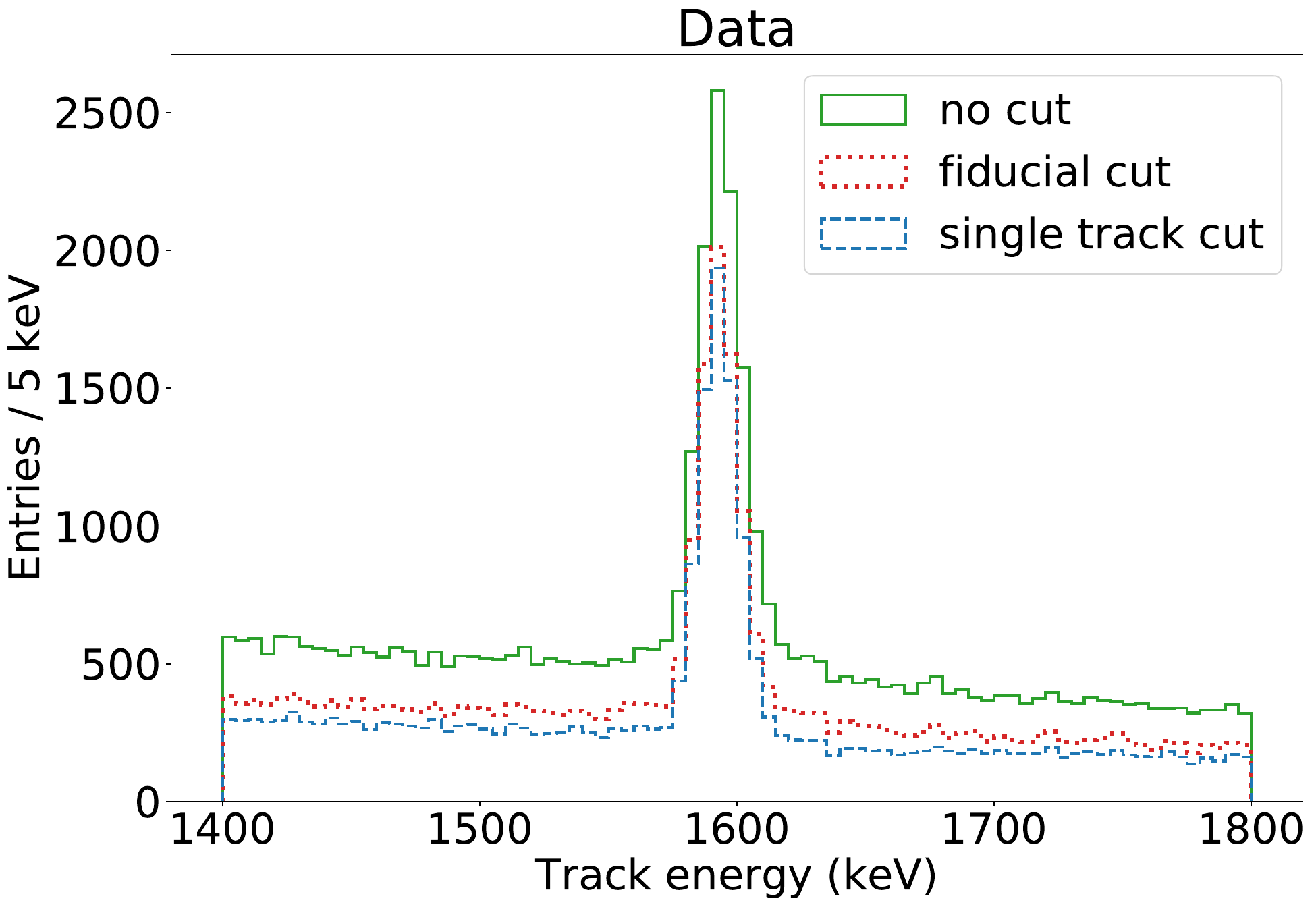}
\caption{\small Distribution of the track energy in the region around the double escape peak before any cut and after the single-track and the fiducial cut (see text for details). Left: Monte Carlo, right: data.}
\label{fig.spectra}
\end{figure}

After the reconstruction step, which provides a set of tracks for each event, a first fiducial filter is applied in order to identify events with the correct energy in the double escape peak. Both data and Monte Carlo samples are required to be fully contained in a fiducial volume, defined as the volume contained within $\simeq$2 cm from all the borders of the drift region, namely $r <$180 mm, and 20 mm $< z <$ 510 mm. A track is considered fully contained if each one of its hits lies inside the fiducial volume.  Subsequently, a second filter is applied, which requires that the events have a single track. This filter is found to clean up the region of the peak by eliminating events  in which a bremsstrahlung or an X ray photon was emitted. It also eliminates tracks with energies in the region of interest that were reconstructed as part of a higher-energy, multi-track event. Removing multi-track events improves the energy resolution of the double escape peak and the modelling of the track energy distribution as a gaussian plus an exponential function, as shown later. 
 In Fig.~\ref{fig.spectra} the track energy distribution is shown before and after these cuts, for Monte Carlo (\textit{left}) and data (\textit{right}). The ratio between the number of events in the peak after the fiducial cut and the initial sample (after subtracting the background statistically) is in agreement between data and Monte Carlo. On the other hand, the single-track filter shows a lower efficiency in Monte Carlo (0.62 $\pm$ 0.01) than in data (0.73 $\pm$ 0.02).  This difference comes from a better reconstruction of tracks in Monte Carlo than in data, which allows one to separate better different energy depositions. However, this difference in the selection does not affect the performance of the topological cut, as shown later. This means that the pattern of the energy deposition at the end-points of the tracks is not affected by reconstructing satellite tracks together with the main one, mainly because the union happens with highest probability in the middle of the track, and not at the extremes.
 
 A last filter is applied during event selection, which ensures that the two blob candidates of a track (defined in the following section) do not overlap. At the energies of the \TL\  double escape peak this requirement has little effect, rejecting less than 2$\%$ of tracks both in data and Monte Carlo.




\section{Topological discrimination}
\label{topo}
\subsection{Blob candidate definition}\label{blobdef}
The aim of this work is to assess the performance of a cut on the energy of the end-points of the track, as a means to discriminate signal from background (the blob cut). For each track, two blob candidate energies are defined by summing the energy of the hits contained in a sphere of fixed radius centred on the end-points previously identified with the BFS algorithm, as explained in Sec.~\ref{reco}.  It can happen that hits are included in the blob candidate that are far away from the extreme in terms of distance measured along the track, but have a short Euclidean distance from it (as, for example, in the case of a winding track). In order to avoid this, only the hits belonging to the voxels that have a distance along the track shorter than the radius plus an allowance are considered. The allowance is needed because the voxel position is discretized, therefore an extra distance equivalent to the size of the voxel diagonal is added to the radius, only to select the voxels, to ensure that all the hits within the spheres are taken correctly into account. Once the voxel candidates are selected, only the hits belonging to those voxels and that have a Euclidean distance shorter than the radius from the end-points of the tracks are considered for the blob candidates. A detailed description of the optimization of the value of the blob radius is given in Appendix \ref{blob_opt}.

\begin{figure}[htb!]
\centering
\includegraphics[width=.45\textwidth]{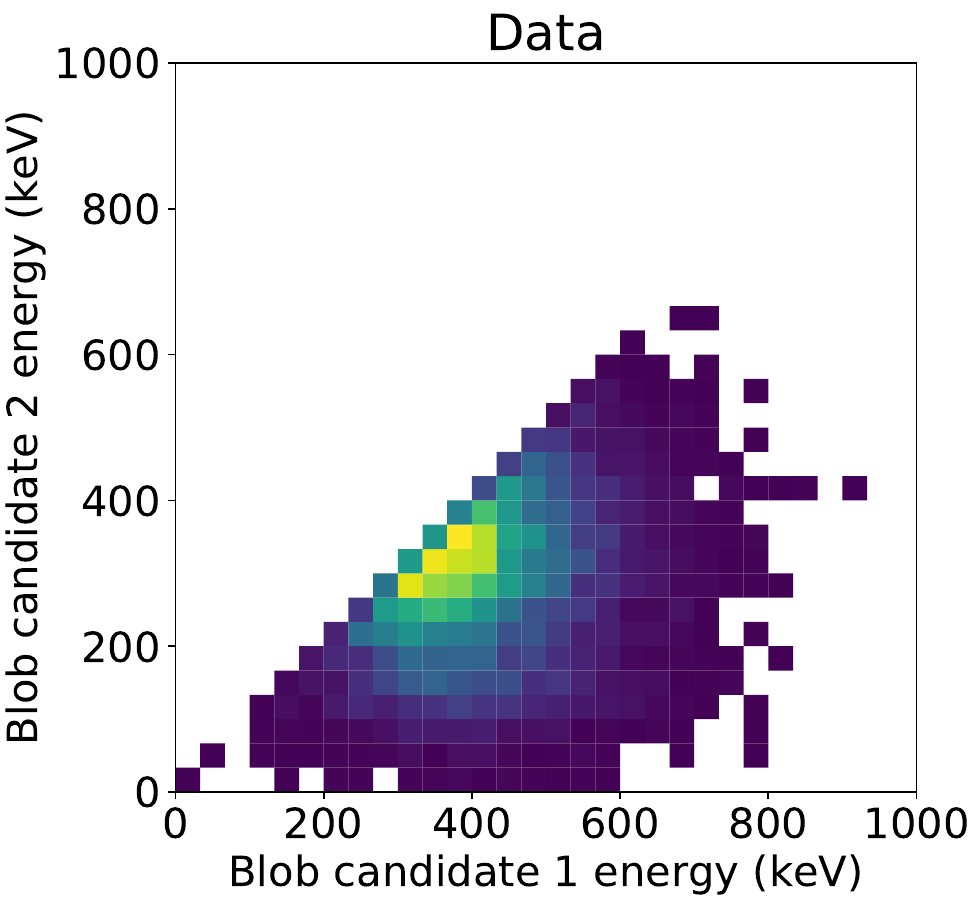}
\includegraphics[width=.515\textwidth]{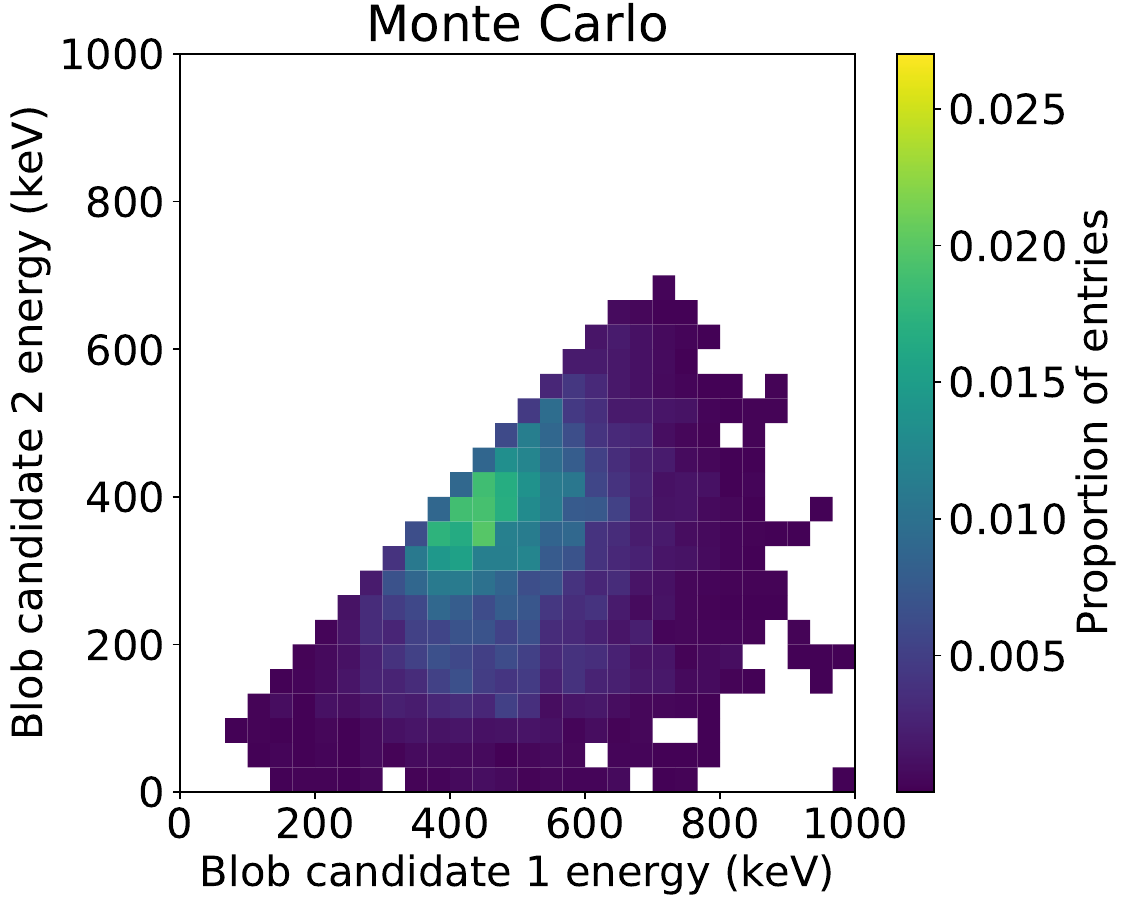}
\caption{\small Distribution of the energies of \blobone\ and \blobtwo\ for data (left) and Monte Carlo (right), in the 1570--1615 keV region, under the \TL\ double escape peak.}
\label{fig.blob_e}
\end{figure}
In Fig.~\ref{fig.blob_e} the energy distribution of the two blob candidates is shown for data and Monte Carlo, for tracks with energies in the \TL\ double-escape peak (1570--1615 keV). On the $x$ and $y$ axis, the energy of the higher energy blob candidate (from now on, \blobone) and the energy of the lower energy blob candidates (\blobtwo) are represented, respectively. A cut on the energy of \blobtwo\ will be applied, to separate background from signal.

\begin{figure}[htb!]
\centering
\includegraphics[width=.49\textwidth]{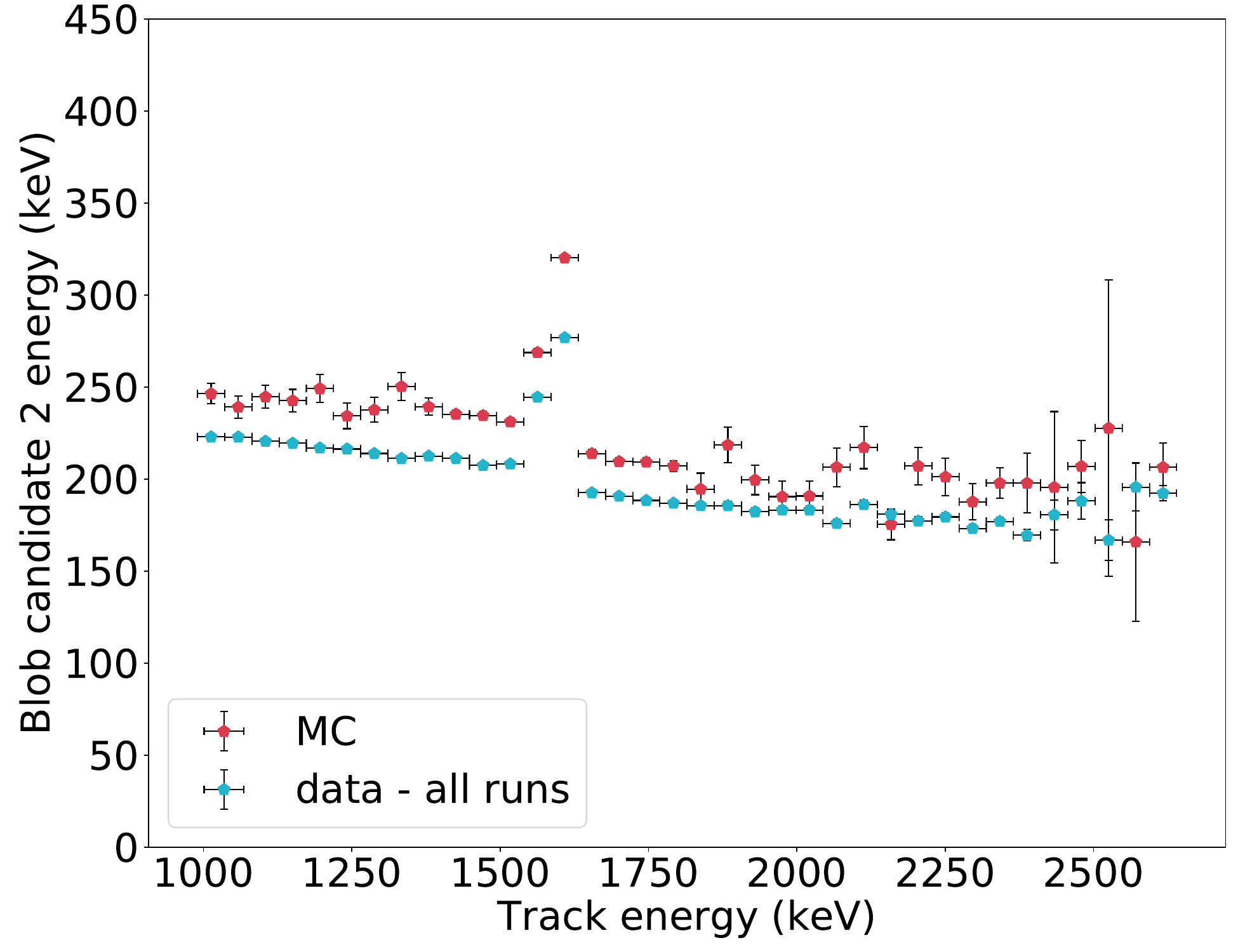}
\includegraphics[width=.49\textwidth]{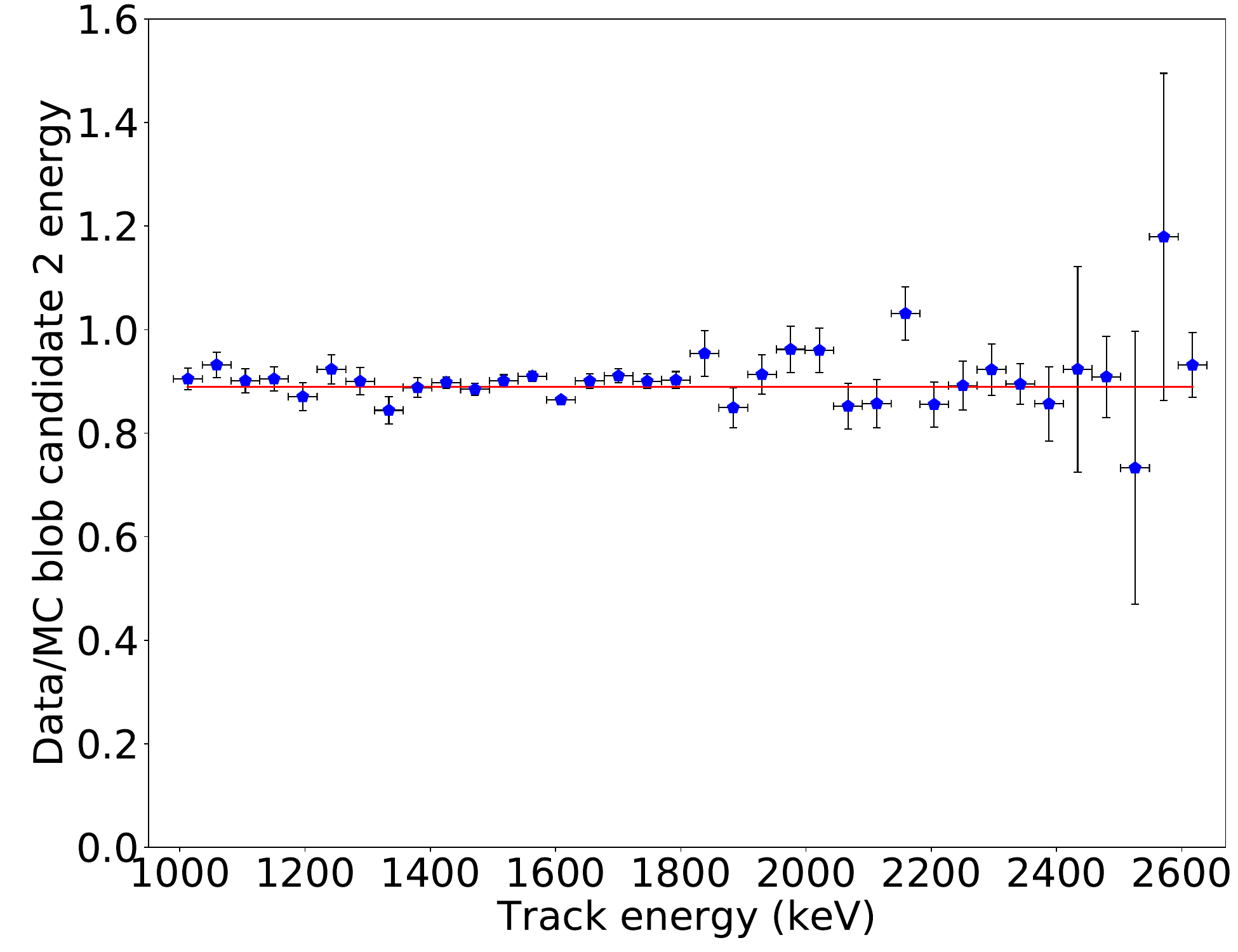}
\caption{\small Energy of \blobtwo\ as a function of the track energy for data and Monte Carlo. The left panel shows the average value for each bin, while the right panel shows the ratio between data and Monte Carlo and a linear fit to the distribution.}
\label{fig.ratio}
\end{figure}
\begin{figure}[htb!]
\centering
\includegraphics[width=.49\textwidth]{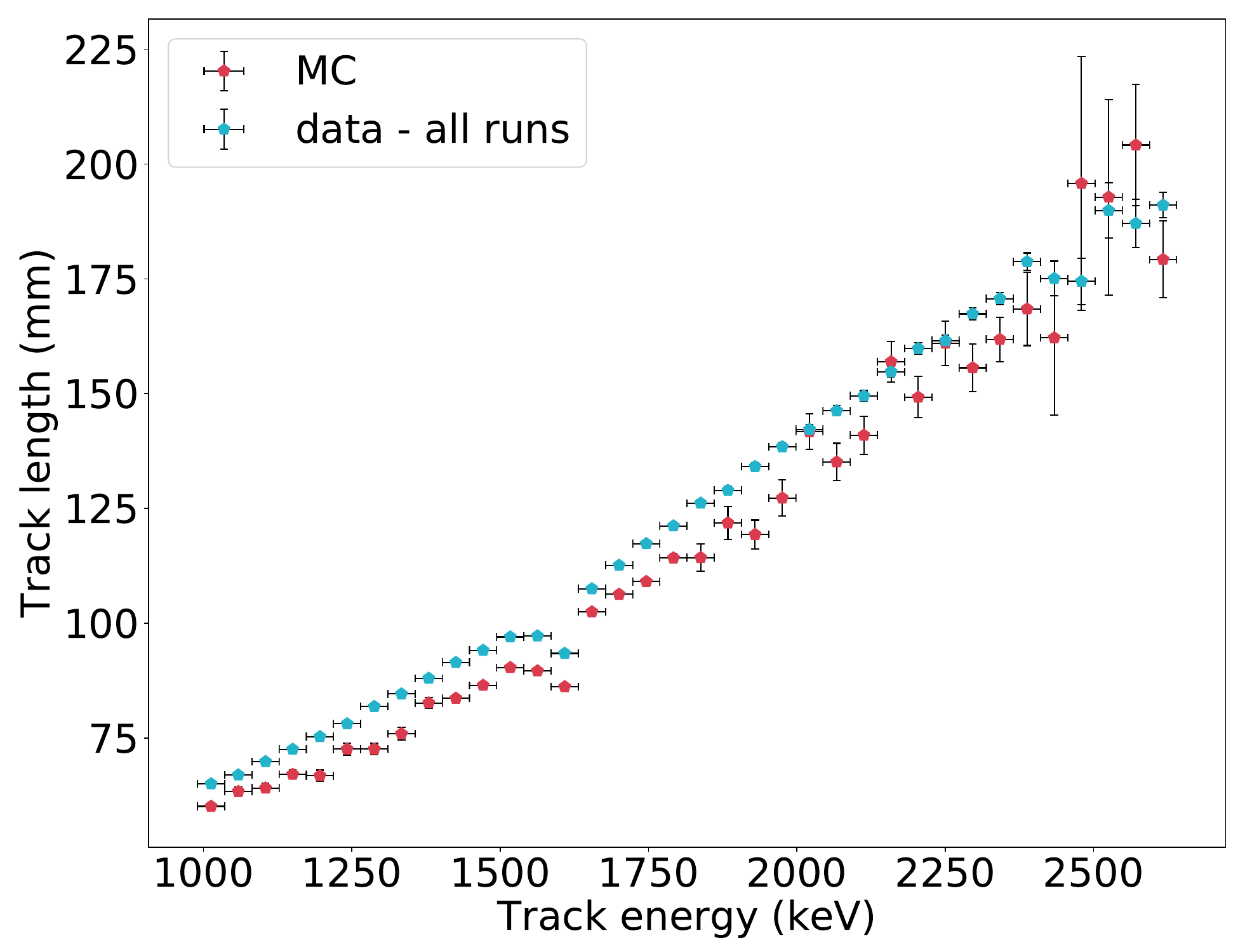}
\includegraphics[width=.49\textwidth]{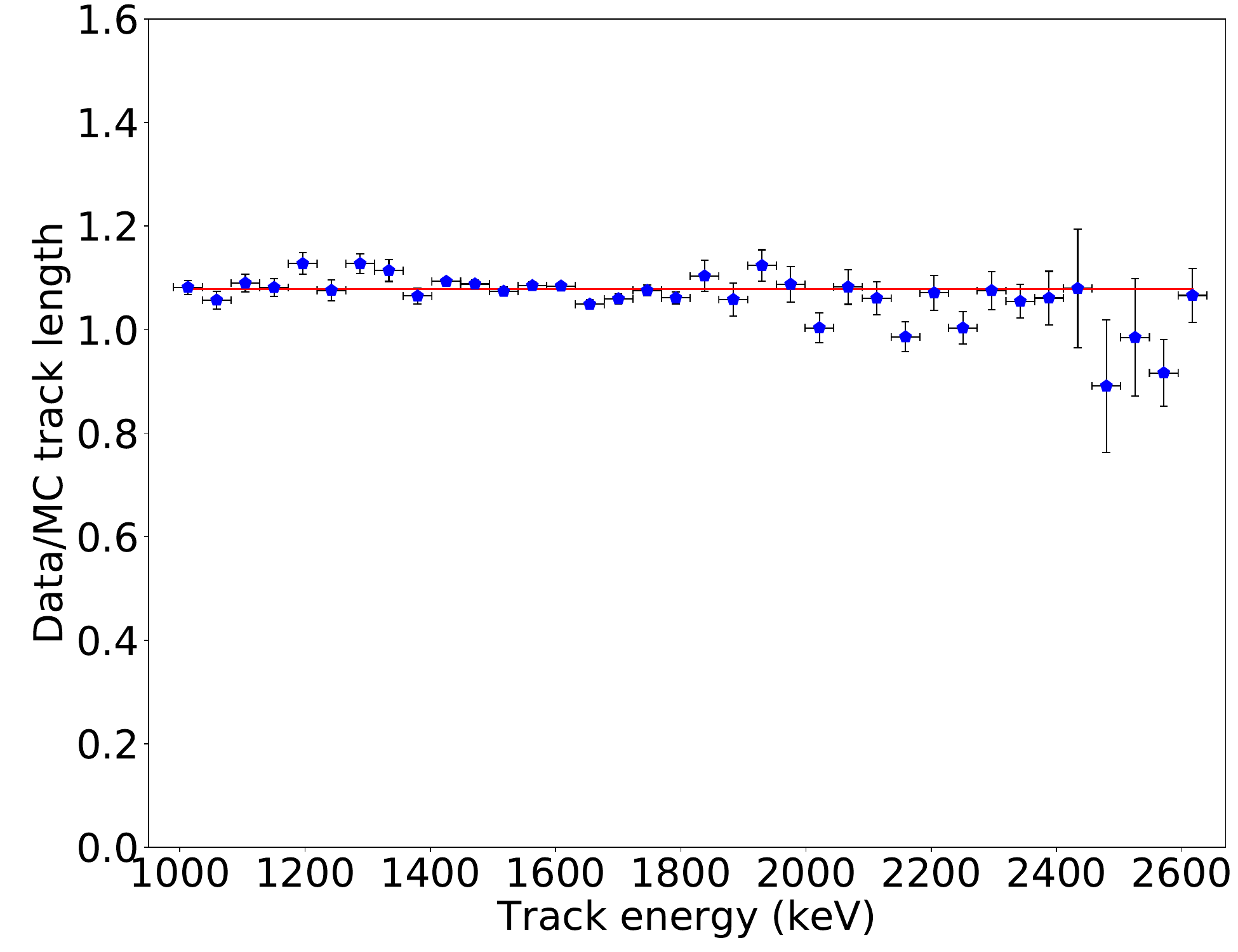}
\caption{\small Track length as a function of the track energy for data and Monte Carlo. The left panel shows the average value for each bin, while the right panel shows the ratio between data and Monte Carlo and a linear fit to the distribution.}
\label{fig.length}
\end{figure}
A difference in the energy of the two blobs appears between data and Monte Carlo, specifically, a difference of around $11\%$ is measured for \blobtwo, which is the observable used in the cut, and Monte Carlo has the higher energy. While \blobone\ is the end-point of the track for all the entries of the sample, \blobtwo\ is the starting point of the track only for a fraction of them, while for events of pair production it represent the second, real blob. The estimated fraction of signal-like and background-like events is similar in data and Monte Carlo, with a prevalence of the former (75.0 $\pm$ 1.1\% vs 25.0 $\pm$ 0.2\% for Monte Carlo and 75.8 $\pm$ 1.1\% versus 24.2 $\pm$ 0.2\% for data). To understand how this discrepancy between data and Monte Carlo behaves on the true starting points of the track, the energy of \blobtwo\ is studied also in energy regions outside the double escape peak, as shown in Fig.~\ref{fig.ratio}\textit{--left}.  The ratio between the \blobtwo\ energy of data and Monte Carlo is roughly constant across the energy spectrum up to the thallium photopeak (Fig.~\ref{fig.ratio}\textit{--right}) and it is fitted to a straight line which provides a best value for the ratio equal to $0.889 \pm 0.003$. The energy bins right before the last one, which corresponds to the photopeak, have very low statistics, thus large error bars, because they are in between the photopeak and the Compton edge of the 2.615-MeV gamma, where very few events are present. The reason of this difference in track end-point energies between data and Monte Carlo is currently under study. It could be related with a difference in the length of the tracks, which appears to be larger in data than in Monte Carlo on average, across the energies, as shown in Fig.~\ref{fig.length}. The inverse parallelism between the data/MC ratio of the two observables (\blobtwo\ energy and the length of a track) is evident, as can be appreciated in the left panel of Fig.~\ref{fig.ratio} and Fig.~\ref{fig.length}. The ratio between the length of data and Monte Carlo tracks can be fitted to a straight line, with best value for the constant term equal to  $1.078 \pm 0.002$ (Fig.~\ref{fig.length}\textit{--right}). Also, the multiple scattering model used in our \textsc{Geant4} detector simulation can affect the energy density distribution along the electron track. A second simulation has been run, changing the \textsc{Geant4} version to the latest one to this date (\texttt{geant4.10.5.p01}) and using the recommended physics list for electrons below 100 MeV (namely, \texttt{G4EmStandardPhysics\_option4}). No significant variations in the energy of the \blobtwo\ is found. 

The Monte Carlo energy of the \blobtwo\  is then rescaled, being multiplied by the ratio extracted from the fit.  The error of the fit and the variation of the results using a different \textsc{Geant4} distribution are included in the calculation of the systematic uncertainties on the signal and background efficiencies.


\subsection{Efficiency calculation}
%
%
\begin{figure}[htb!]
\centering
\includegraphics[width=.49\textwidth]{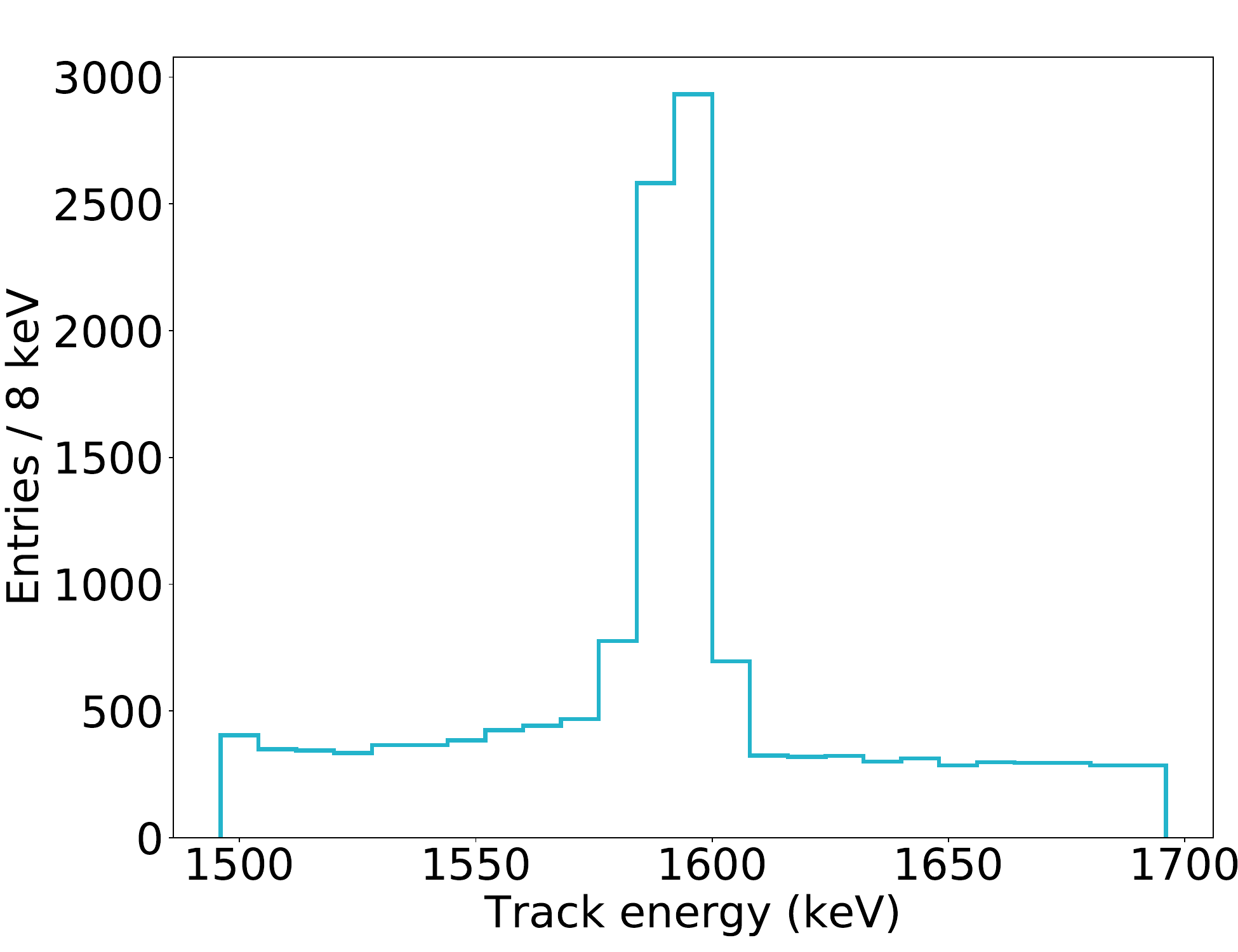}
\includegraphics[width=.49\textwidth]{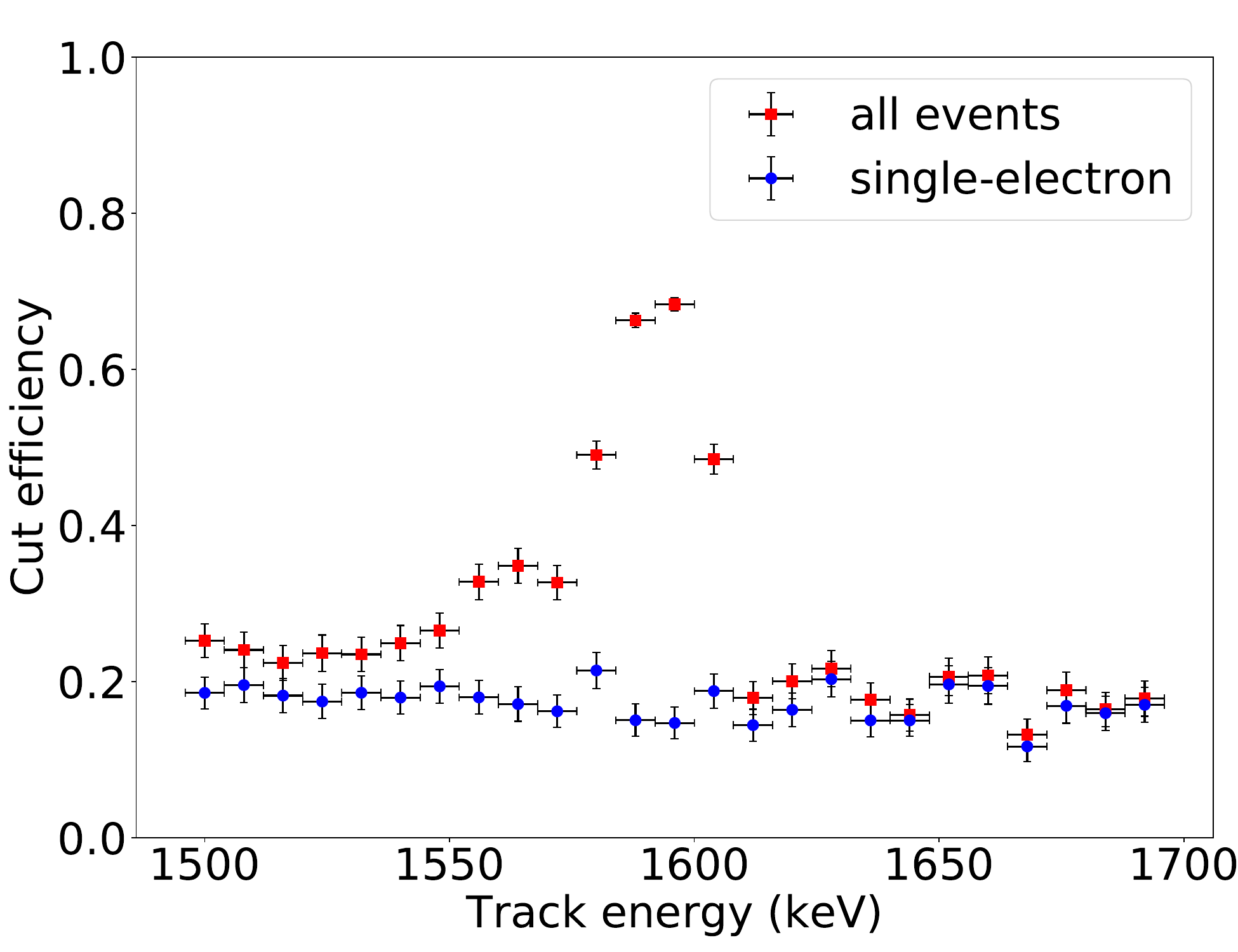}
\caption{\small Blob cut efficiency for Monte Carlo, using the full event sample and a sample of single-electron events. The histogram on the left side represents the full sample of track energies and the plot on the right side shows the fraction of the total number of events passing the cuts for each one of the two samples.}
\label{fig.cut}
\end{figure}

In Fig.~\ref{fig.cut} the efficiency of the blob cut as a function of the track energy is shown for the full sample of selected events and for a sample where only background-like events are retained, where this selection is made using Monte Carlo truth. The plot illustrates the fact that in the energy bins where a mixture of signal-like and background-like events is present, the total efficiency of the cut increases dramatically, while the efficiency of the background-like events stays constant. It is worth to notice that the energy bins close to the left side of the double escape peak are populated by those e+e-- events that have emitted bremsstrahlung gammas, losing part of their energy; this explains the higher efficiency compared to the right side of the peak. 

To calculate the efficiency of the blob cut on double electron tracks, we need to identify a sample of pure signal-like and background-like events. Since the \TL\ double escape peak region is populated by both electron-positron pairs and Compton electrons, a fit to a gaussian+exponential function is applied to the track energy spectrum of the events that pass the selection described in Sec.~\ref{sec.ev.sel}, to separate the two samples statistically. In fact, the electron-positron track energies are expected to be gaussianly distributed, with a mean at $1593$ keV, which is the energy of the 2.615-MeV de-excitation gamma of \TL\ minus the energy of the two 511-keV gammas originating from the positron annihilation. On the other hand, the single-electron tracks come from the Compton continuum of the 2.615-MeV gamma, which can be modelled as an exponential function.
\begin{figure}[htb!]
\centering
\includegraphics[width=.49\textwidth]{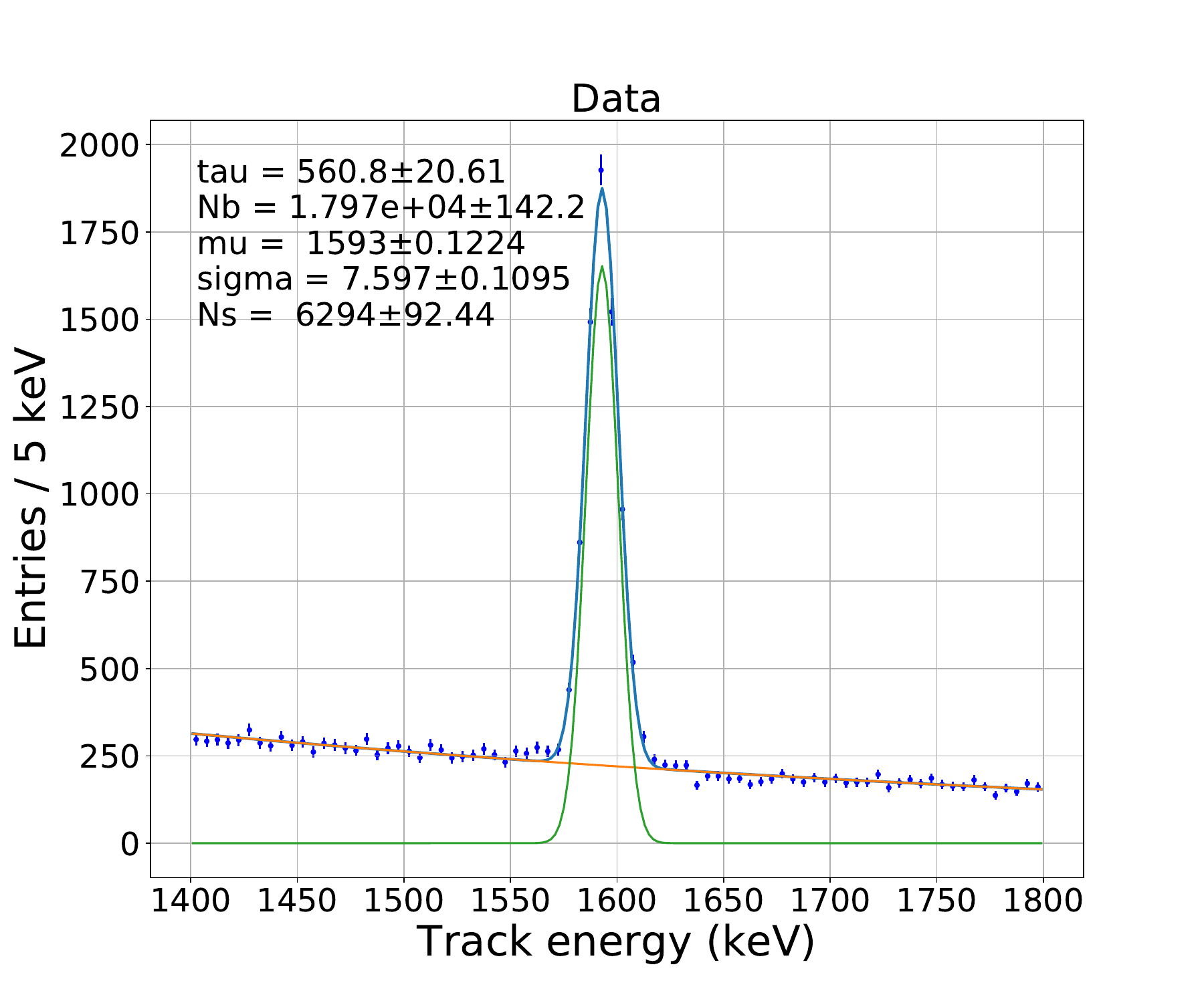}
\includegraphics[width=.49\textwidth]{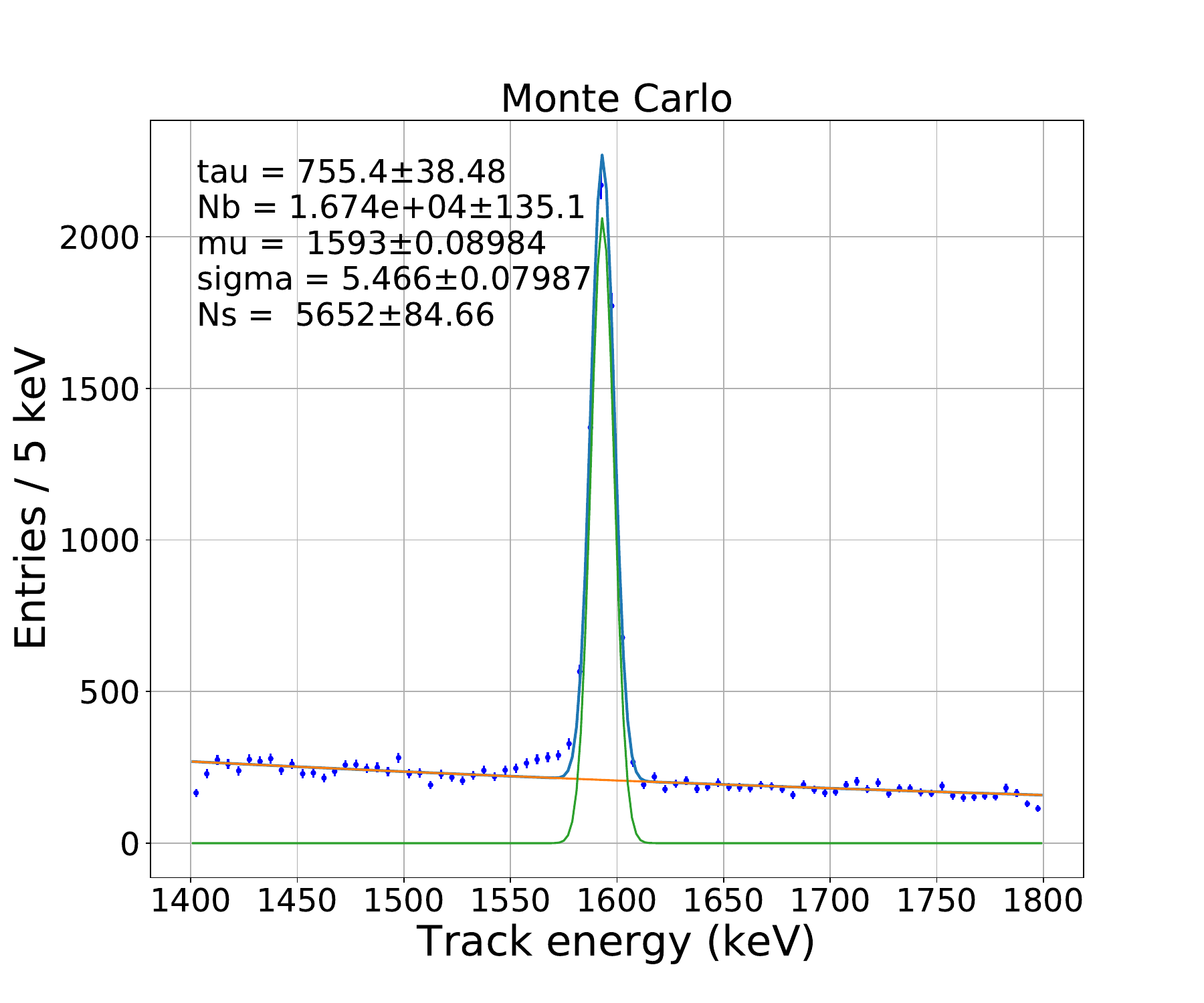}
\caption{\small Energy spectrum and results of the gaussian+exponential fit for both data (left) and Monte Carlo (right).}
\label{fig.fit}
\end{figure}

An unbinned maximum likelihood fit is applied to the track energy spectrum, in the region between 1400 and 1800 keV, and the number of signal-like and background-like events in the double-escape peak region is calculated integrating the gaussian and exponential functions evaluated with the parameters obtained by the fit, in a pre-defined range between 1570 and 1615 keV. This range is large enough to contain virtually the whole gaussian peak (more than 99.5 \% of the area) for both data and Monte Carlo. The dependence of the results on the chosen ranges has been accounted for in the systematic error. The result of the fit is shown in Fig.~\ref{fig.fit}. A variable threshold is applied on the energy of the two blob candidates, starting from 0 up to 500 keV. After each cut, the number of signal and background events is recalculated performing the fit on the energies of the tracks that pass the cut. The cut efficiency for both signal-like and background-like events is given by the ratio between the number of events of each type after the cut and the initial number of events of that type.
\begin{figure}[htb!]
\centering
\includegraphics[width=0.49\textwidth]{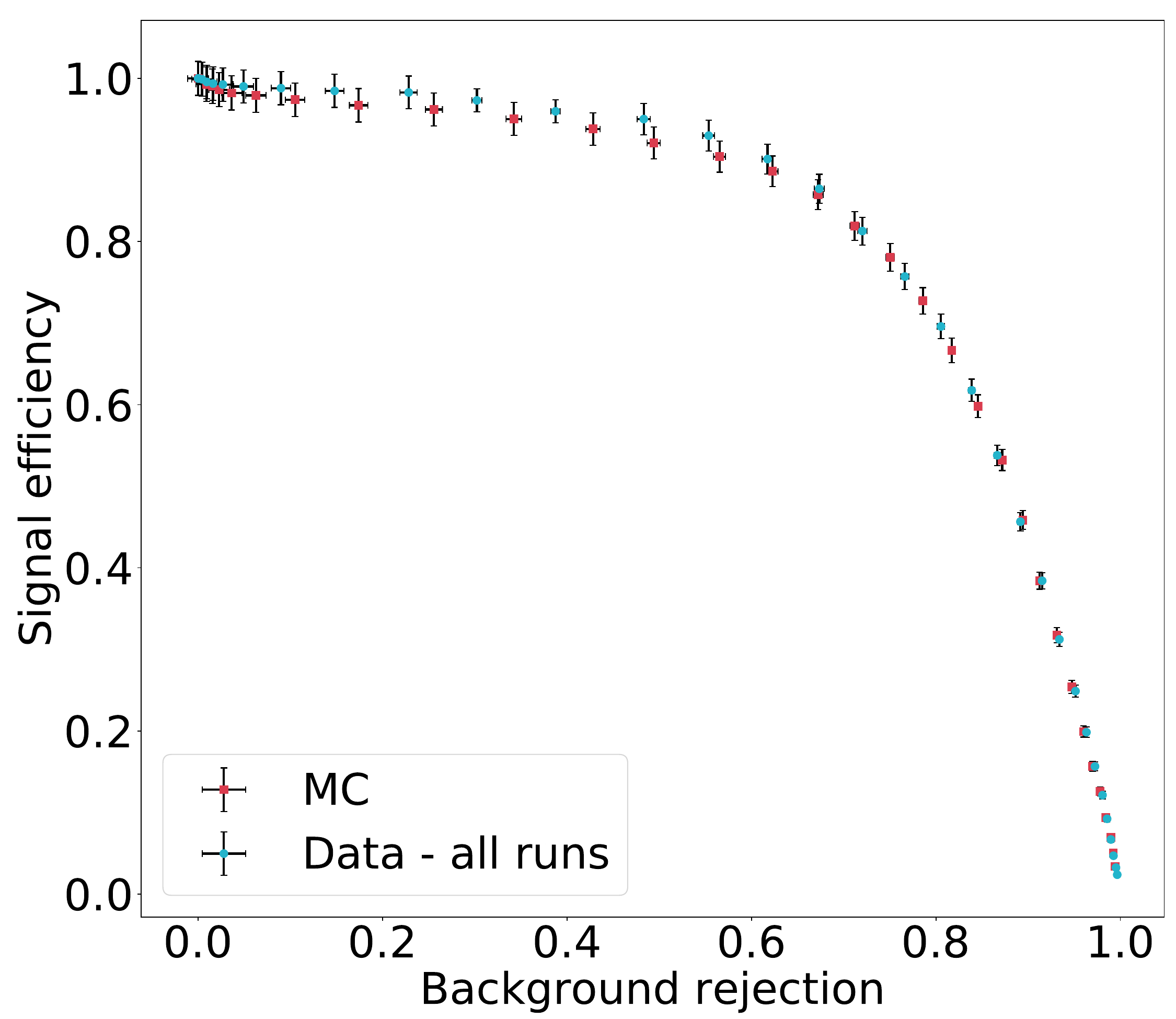}
\includegraphics[width=0.49\textwidth]{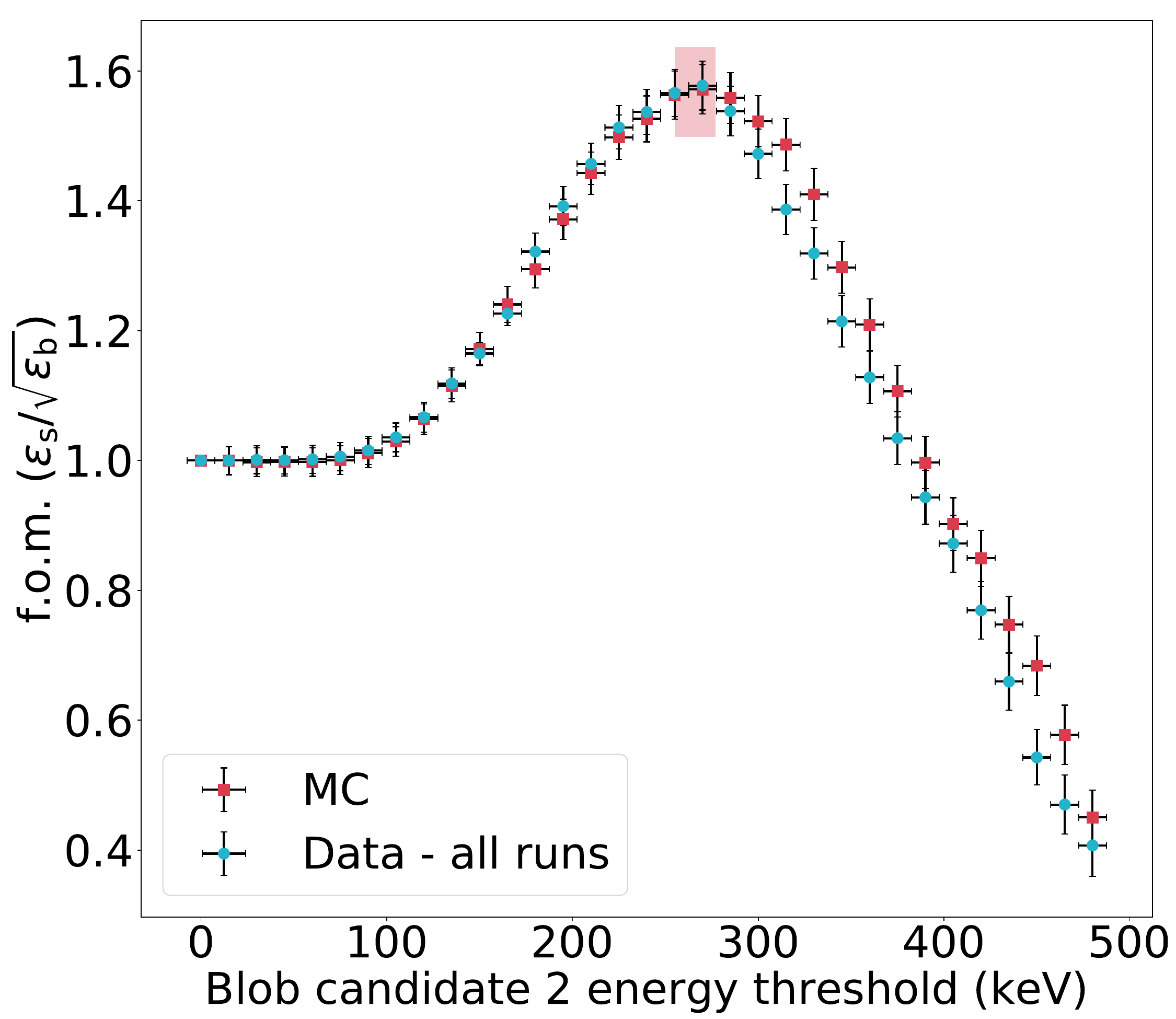}
\caption{\small Left: Signal efficiency as a function of background rejection (proportion of background events removed from the sample by the blob cut), varying the required minimum energy of the \blobtwo. Right: Figure of merit (defined in Eq.~\ref{eq.fom}) as a function of the threshold on the energy of the \blobtwo\ after rescaling Monte Carlo. The highlighted area corresponds to the best threshold. In both figures, data and Monte Carlo simulation are shown.}
\label{fig.roc}
\end{figure}
In Fig.~\ref{fig.roc}\textit{-left} the signal efficiency and the background rejection (defined as the fraction of background events that do not pass the cut) are plotted for each value of the threshold, for both data and Monte Carlo, showing very good agreement.

In order to choose the best value for the threshold, the following figure of merit is maximized:
\begin{equation}
\frac{\varepsilon}{\sqrt{b}}\,
\label{eq.fom}
\end{equation}
where $\varepsilon$ and $b$ are the fraction of signal events and the fraction of background events passing the cut, respectively. This quantity is an estimator of the discrimination power of the topological cut, since the sensitivity to the half-life of the \bbonu\ decay is directly proportional to the detector efficiency and inversely proportional to the square root of the rate of background in background-limited experiments \cite{GomezCadenas:2010gs}. In Fig.~\ref{fig.roc}\textit{--right}, this figure of merit is displayed as a function of the threshold, for data and Monte Carlo. The best value of the threshold is then calculated taking the mean of the values of the threshold around that of the maximum figure of merit, in an interval for which the figure of merit is between 99\% of the maximum and the maximum.



\section{Discussion}
The value of the \blobtwo\ energy threshold that optimizes the performance of the blob cut in data is $265.9 \pm 0.6_{\textrm{ sys}}$ keV and the efficiency obtained for pure signal-like events is $71.6 \pm 1.5_{\textrm{ stat}} \pm 0.3_{\textrm{ sys}} \%$ for a background acceptance of $20.6 \pm 0.4_{\textrm{ stat}} \pm 0.3_{\textrm{ sys}} \%$. The same cut applied to Monte Carlo data gives a signal efficiency of $73.4 \pm 1.2 _{\textrm{ stat}} \pm 3.0_{\textrm{ sys}} \%$ for a background acceptance of $22.3 \pm 0.4_{\textrm{ stat}} \pm 0.5_{\textrm{ sys}} \%$, in agreement with data. This result, which corresponds to a figure of merit of $1.578 \pm 0.038_{\textrm{ stat}}  \pm 0.005_{\textrm{ sys}}$,  is an improvement of the topological discrimination compared to the measurement carried out in the NEXT-DEMO prototype, where a figure of merit of 1.35 was reached. This improvement is due to the larger dimensions of the NEXT-White detector, which allows for a better reconstruction of longer tracks, where the two end-points are well separated.

\begin{figure}[htb!]
\centering
\includegraphics[width=.7\textwidth]{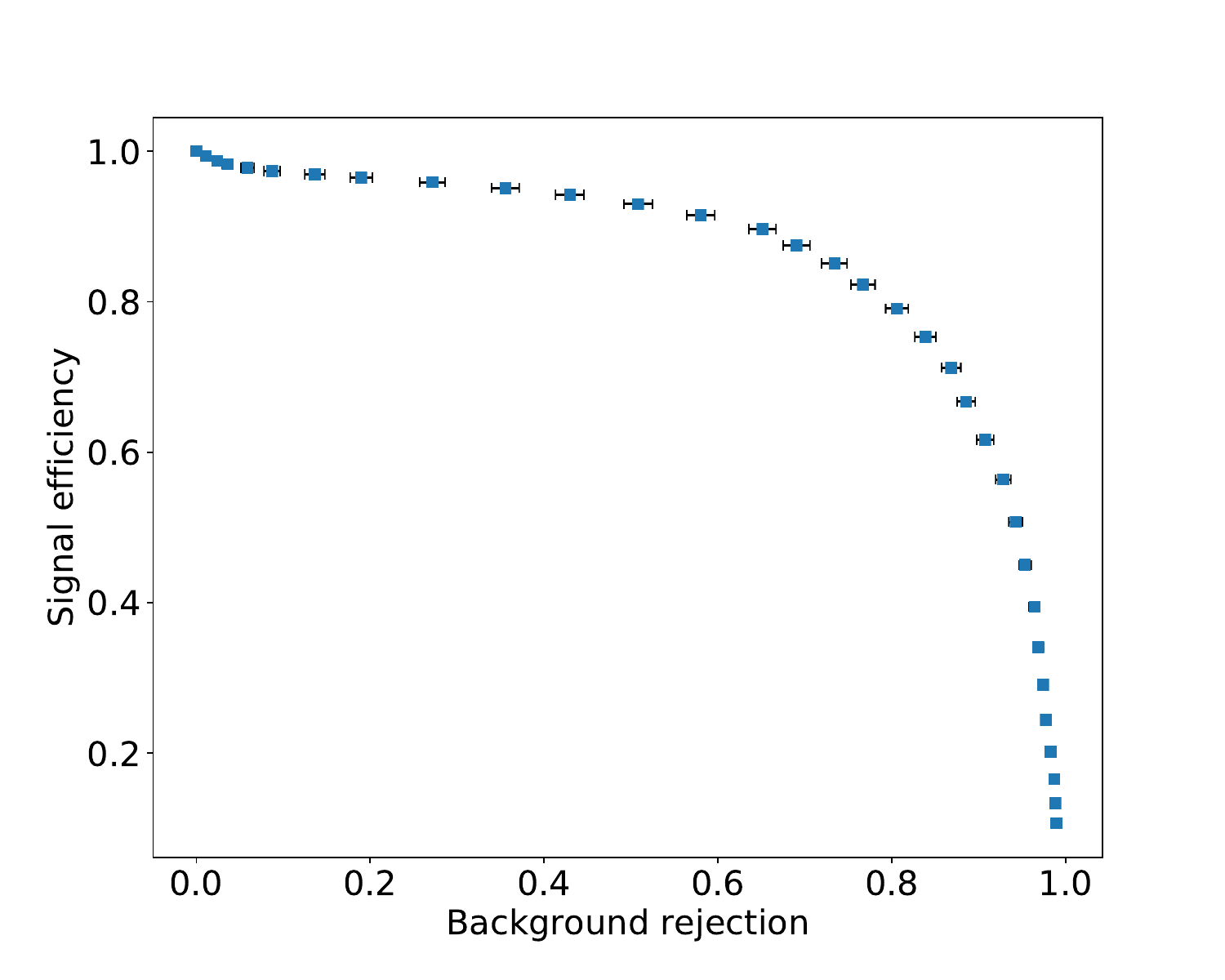}
\caption{\small Signal efficiency as a function of background rejection, varying the required minimum energy of the \blobtwo\ between 0 and 500 keV, in the \bbonu\ region.}
\label{fig.0nu}
\end{figure}

Having tuned our Monte Carlo model and then demonstrated the good agreement between data and Monte Carlo in the \TL\ double escape peak region, it is possible to study the efficiency of the blob cut in the \bbonu\ region, with Monte Carlo simulations, and extrapolate the results to data. With this aim, two dedicated samples have been simulated, with large statistics, with the same detector conditions as the \TL\ calibration source sample used in the double escape peak analysis. The first one is a sample of \bbonu\ decays of \XE, uniformly distributed in the active volume, while the second one is generated from nuclei of \TL\ distributed in the teflon light tube that surrounds the active volume, which is one of the dominant contributions in the NEXT-White background model \cite{Novella:bckg}. If the 2.615-MeV thallium de-excitation gamma produced in the decay converts in the xenon through photoelectric interaction, the resulting photoelectron can lose part of its energy via bremsstrahlung radiation, therefore the energy of its track can fall in the region around the \XE\ Q-value.

The same reconstruction, selection and analysis used for the \TL\ double escape peak region are applied, within an energy window of around 45 keV around the \bbonu\ peak, (namely, 2435--2481 keV) and the curve of signal efficiency versus background rejection for the \bbonu\ region is shown in Fig.~\ref{fig.0nu}.  The samples are rescaled by the same factor applied to the double escape peak analysis, which was found to be constant across energies.
A threshold of 266.5 keV gives a signal efficiency of $71.5 \pm 0.1_{\textrm{ stat}} \pm 0.3_{\textrm{ sys}} \%$, for a background acceptance of $13.6 \pm 1.1_{\textrm{ stat}} \pm 0.7_{\textrm{ sys}} \%$. The blob cut appears to cut more background at the \bbonu\ energies than at the lower energies of the \TL\ double escape peak. This improvement in the performance of the blob cut is due to the fact that the tracks are larger, and therefore the separation between their end-points is better defined. However, a larger improvement is expected to be reached at the \bbonu\ energy, with modifications in the reconstruction process.  For instance, Fig.~\ref{fig.blob_radius}\textit{--right} indicates that a larger radius would benefit, without loosing events that present blob overlap since high energy tracks are longer. Furthermore, different algorithms based, for instance, on Deep Neural Networks or Maximum-Likelihood Expectation Maximization have the potential to improve the topological discrimination \cite{Renner:2016trj, Simon:2017pck}. The discrimination efficiency is expected to be increased also by having a gas with reduced electron transverse diffusion, such as Xe with subpercent concentration of a molecular additive \cite{Henriques:2017rlj, Henriques:2018tam} or Xe-He mixtures \cite{Felkai:2017oeq, McDonald:2019fhy}.


\label{res}

\section{Conclusions}
\label{conclus}
In this work, the power of the topological discrimination of signal from background has been explored in the NEXT-White detector. Electron-positron pair tracks have been used to mimic the \bbonu\ signal, while single-electron tracks coming from Compton interactions, at the same energy, have been used as a background sample. The difference in the deposited energy at the beginning and at the end of an electron (or positron) track has been exploited to define a cut to separate signal from background, namely, a threshold on the lower energy extreme of a track. A threshold of $265.9 \pm 0.6_{\textrm{ sys}}$ keV provides a signal efficiency of $71.6 \pm 1.5_{\textrm{ stat}} \pm 0.3_{\textrm{ sys}} \%$ for a background acceptance of $20.6 \pm 0.4_{\textrm{ stat}} \pm 0.3_{\textrm{ sys}} \%$. This result improves on the one reported in Ref.~\cite{Ferrario:2015kta}, thanks to an improved track reconstruction, and also due to the larger dimensions of the detector. 

The agreement of the blob cut performance between data and Monte Carlo simulation after calibrating the \blobtwo\ energy is good, therefore a study of the same cut has been carried out with Monte Carlo to estimate the expected performance in the \bbonu\ energy region. A signal efficiency of $71.5 \pm 0.1_{\textrm{ stat}} \pm 0.3_{\textrm{ sys}} \%$ is predicted, for a background acceptance of $13.6 \pm 1.1_{\textrm{ stat}} \pm 0.7_{\textrm{ sys}} \%$.  The background rejection improves at higher energy, due to the fact that electron tracks are on average larger, therefore the blobs are better defined. The background electron sample used in this analysis comes from \TL. Another source of background for \bbonu\ searches comes from the 2.448-MeV gamma of \BI\ decay. Detailed Monte Carlo studies have demonstrated that similar rejection factors are expected in this case, see for example \cite{Martin-Albo:2015rhw, javithesis}.


In summary, we have proven from the data themselves the power of the NEXT technology to separate the two-electron signal characteristic of double beta decays from single-electrons emanating from backgrounds. Our current result improves the first study carried out with the NEXT-DEMO prototype and allows for a robust extrapolation to the \Qbb\ energy, confirming the excellent performance of the unique topological signature of NEXT.


\acknowledgments
The NEXT Collaboration acknowledges support from the following agencies and institutions: the European Research Council (ERC) under the Advanced Grant 339787-NEXT; the European Union's Framework Programme for Research and Innovation Horizon 2020 (2014-2020) under the Marie Sk\l{}odowska-Curie Grant Agreements No. 674896, 690575 and 740055; the Ministerio de Econom\'ia y Competitividad and the Ministerio de Ciencia, Innovaci\'on y Universidades of Spain under grants FIS2014-53371-C04, RTI2018-095979, the Severo Ochoa Program SEV-2014-0398 and the Mar\'ia de Maetzu Program MDM-2016-0692; the GVA of Spain under grants PROMETEO/2016/120 and SEJI/2017/011; the Portuguese FCT under project PTDC/FIS-NUC/2525/2014, under project UID/FIS/04559/2013 to fund the activities of LIBPhys, and under grants PD/BD/105921/2014, SFRH/BPD/109180/2015 and SFRH/BPD/76842/2011; the U.S.\ Department of Energy under contracts number DE-AC02-06CH11357 (Argonne National Laboratory), DE-AC02-07CH11359 (Fermi National Accelerator Laboratory), DE-FG02-13ER42020 (Texas A\&M) and DE-SC0019223 / DE-SC0019054 (University of Texas at Arlington); and the University of Texas at Arlington. DGD acknowledges Ramon y Cajal program (Spain) under contract number RYC-2015-18820. We also warmly acknowledge the Laboratori Nazionali del Gran Sasso (LNGS) and the Dark Side collaboration for their help with TPB coating of various parts of the NEXT-White TPC. Finally, we are grateful to the Laboratorio Subterr\'aneo de Canfranc for hosting and supporting the NEXT experiment.

\appendix
\section{Blob cut optimization}\label{blob_opt}

\begin{figure}[htb!]
\centering
\includegraphics[width=.49\textwidth]{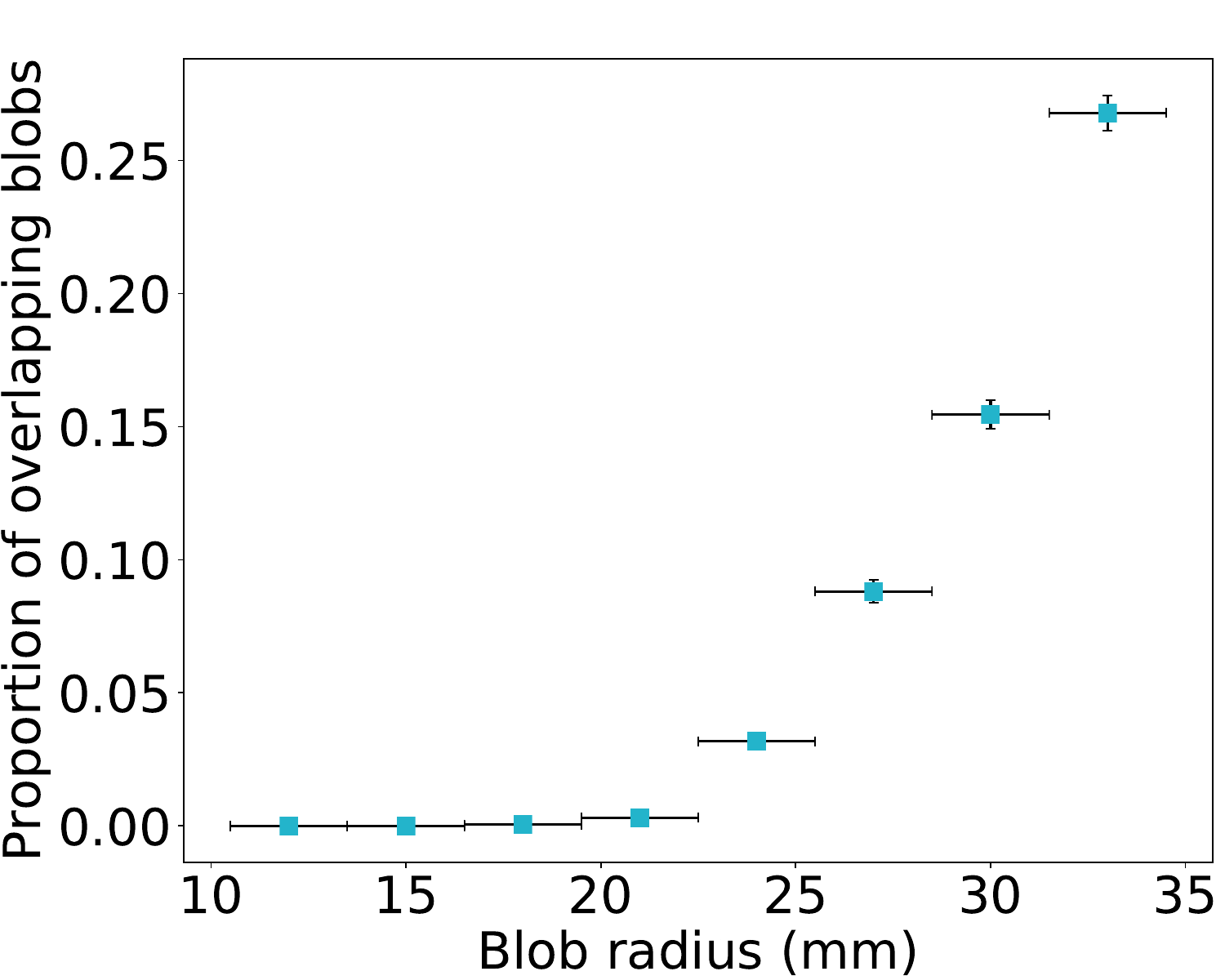}
\includegraphics[width=.49\textwidth]{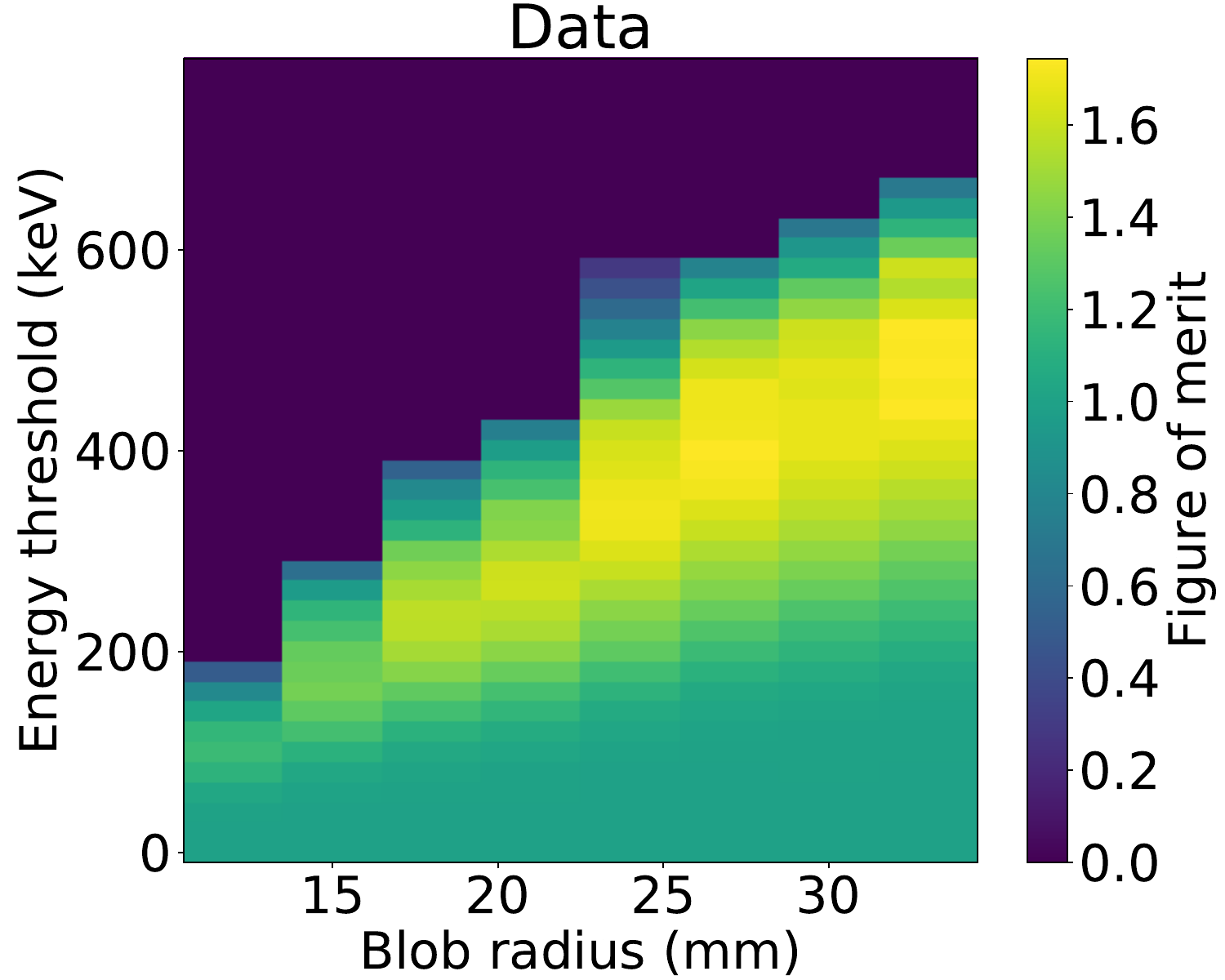}
\caption{\small Proportion of events with overlapping blob candidates (left) and dependence of the figure of merit on the blob radius and the energy threshold on the \blobtwo\ (right).}
\label{fig.blob_radius}
\end{figure}
Several values for the blob candidate radius have been considered with the aim of maximizing the discrimination power of the topological cut and at the same time keeping the percentage of tracks with the two blob candidates overlapping to a minimum. The figure of merit used for this optimization is the same as in Eq.\eqref{eq.fom}. In Fig.~\ref{fig.blob_radius}\textit{-left} the fraction of tracks that present overlapping blob candidates is shown for different radii, for events in the double escape peak, while in Fig.~\ref{fig.blob_radius}\textit{-right} the figure of merit is shown as a function of the blob candidate radius and the value of the energy threshold on the \blobtwo. A radius of 21 mm is chosen, since it provides a figure of merit among the highest ones and keeps the fraction of blob overlaps below 2\%.  

\bibliographystyle{JHEP}
\bibliography{Refs}

\end{document}